\documentclass[11pt]{article}
\usepackage{graphicx,caption,subfigure}
\makeatletter
\newcommand{\singlespacing}{\let\CS=\@currsize\renewcommand{\baselinestretch}{1.0}\tiny\CS}
\newcommand{\doublespacing}{\let\CS=\@currsize\renewcommand{\baselinestretch}{1.5}\tiny\CS}

\begin{document}
\title{Analysing $J/\Psi$ Production in Various RHIC Interactions with a Version of Sequential Chain Model (SCM) }
\author{P. Guptaroy$^1$\thanks{e-mail:
gpradeepta@rediffmail.com}, Tarun K. Garain$^2$, Goutam Sau$^3$\thanks{e-mail: gautamsau@yahoo.co.in},
\\
 S. K.
Biswas$^4$\thanks{e-mail: sunil$\_$biswas2004@yahoo.com}, $\&$ S. Bhattacharyya$^5$\thanks{e-mail:
bsubrata@isical.ac.in
(Communicating Author).}\\
{\small $^1$ Department of Physics, Raghunathpur College,}\\
 {\small Raghunathpur 723133, Purulia, West Bengal, India.}\\
 {\small $^2$ Department of Mathematics, Raghunathpur College,}\\
 {\small Raghunathpur 723133, Purulia, West Bengal, India.}\\
 {\small $^3$ Beramara Ram Chandrapur High School,}\\
 {\small South 24-Pgs,743609(WB),India.}\\
{\small $^4$ West Kodalia Adarsha Siksha Sadan, New Barrackpore,}\\
 {\small Kolkata-700131, India.}\\
{\small $^5$ Physics
and Applied Mathematics Unit (PAMU),}\\
 {\small Indian Statistical Institute, Kolkata - 700108, India.}}
\date{}
\maketitle
\bigskip
\bigskip
\begin{abstract}
We have attempted to develop here tentatively a model for $J/\Psi$
production in $p+p$, $d+Au$, $Cu+Cu$ and $Au+Au$ collisions at RHIC
energies on the basic ansatz that the results of nucleus-nucleus
collisions could be arrived at from the nucleon-nucleon
($p+p$)-interactions with induction of some additional specific
features of high energy nuclear collisions. Based on the proposed
new and somewhat unfamiliar model, we have tried (i) to capture the
properties of invariant $p_T$-spectra for $J/\Psi$ meson production;
(ii) to study the nature of centrality dependence of the
$p_T$-spectra; (iii) to understand the rapidity distributions; (iv)
to obtain the characteristics of the average transverse momentum
$<p_T>$ and the values of $<p_T^2>$ as well and (v) to trace the
nature of nuclear modification factor. The alternative approach
adopted here describes the data-sets on the above-mentioned various
observables in a fairly satisfactory manner. And, finally, the
nature of $J/\Psi$-production at Large Hadron Collider(LHC)-energies
deduced on the basis of our chosen model has been presented in a
predictive way against the RHIC-yields, both calculated for the most
central collisions and on the same model.
\end{abstract}

\newpage
\doublespacing
{\flushright{``Finally $J/\Psi$ suppression, which for more than 20 years has
represented the gold-plated signature of deconfinement, is not
understood" - M. J. Tannenbaum \cite{tannenbaum}.}}
\section{Introduction and Background}
At the very start let us first address two somewhat imposed but
important and interesting questions: (i) why are we so seriously
interested in persuing the $\Psi$-studies at both RHIC and LHC?;
(ii) Why do we turn to a somewhat new model, leaving aside the
several versions of the so-called standard model? In fact, the
much-valued observation contained in and captivated by the quotation
at the top from a work by Tannenbaum \cite{tannenbaum} answers, in
part, simultaneously both the questions. But, an elaboration on the
background and the perspective, we feel, is quite necessary for
better understanding and comprehension of both the points to the
desired degree and depth. Why has the $J/\Psi$-suppression attracted
the most attention as the likely ``gold-plated" signal
\cite{bhalerao}?
 The pioneering work of Matsui and Satz\cite{matsui}
established at the very beginning a few points: (i) as a hard QCD
process, the heavy charm pair production takes place very easily,
(ii) the Debye screening of the QGP prevents formation of a $J/\Psi$
state in heavy ion collisions, (iii) the low temperatures at the
hadronization do not really permit of the proposed charm anticharm
pair kinematically; and quite soon after them, it was further
proposed that the suppression pattern ought to have a characteristic
transverse momentum dependence. Later, perturbative calculations
based on perturbative Quantum Chromodynamics (pQCD) established the
fact that both the suppression signal itself as well as  its
$p_T$-dependence could be mimicked by the mundane nuclear shadowing.
It was thus quite evident from the early days that a detailed
quantitative analysis would be necessary to disentangle the effects
of Debye screening in QGP. It has since been recognized that other
effects, notably the simple absorption of the produced $J/\Psi$ in
the nucleus, cause suppression of the produced $J/\Psi$ in all
nucleus-involved collisions, categorized physically as {\bf{normal}}
suppression. The additional or so-called `{\bf{anomalous}}'
suppression as the possible signal of QGP, shall have to be
investigated into only after accommodating the normal suppression.
But, the estimation of these proposed normal/anomalous suppressions
got absolutely entangled in multiple problems in pQCD due to lack of
reliable and consistent knowledge about the several related factors.
The reasons are explained quite well by Bhalerao and Gavai
\cite{bhalerao}. Besides, a detailed review made by Rapp et
al.\cite{rapp08} has pinpointed and elucidated the several
conceptual, difficulties and experimental hurdles encounterde by the
Standard Model (SM) in understanding and interpreting the various
aspects of the $J/\Psi$-data. So, in order to escape these problems
and to minimise the difficulties due to uncertainties, we have tried
here an alternative methodology for understanding the totality or a
large of part of the data on $J/\Psi$-production in some ultrahigh
energy collisions like $p+p$, $d+Au$, $Cu+Cu$ and $Au+Au$
interactions which comprise the totality of the contour of the
experimental RHIC studies \cite{bravina}- \cite{kharzeev08}. The
present study aims at probing and comprehending some of the major
properties and data-characteristics of $J/\Psi$-production in all
the aforementioned RHIC-studies in a model-based manner. And the
model of our choice is named to be the Sequential Chain Model (SCM)
which has no QCD-tag and with some ancillary
 physical ideas. Very initially, and even now, the goal of high energy heavy ion physics has been to study
 `quantum chromodynamics'(QCD) in a regime of high temperature, high density and large reactive volume.
 The hope was and is to find conclusive evidence that `QCD' undergoes a phase transition at a critical temperature
 from a confined state, where so-called quarks and gluons are bound in colourless hadron states, to a hypothetical
 deconfined state named `quark-gluon plasma' (QGP), where `quarks' and `gluons' can explore volumes larger than the
 typical hadron radius ($r\sim1fm$). But in our endeavours made here we would try to avoid using all these standard
 phrases, propositions and methods; rather we would try to build up an entirely new and alternative approach to
 understand the production-mechanism for $J/\Psi$ production and interpret the relevant observables in the
 light of an alternative approach outlined in the next section.
\par
In one of our previous works \cite{pgr05}, we demonstrated in no
uncertain terms that the production of $J/\Psi$ particles was
neither `anomalously' suppressed, nor enhanced; rather it was ``just
normal" where the term `normal' is to be interpreted as the
similarity of the $J/\Psi$-behaviour with production characteristics
of some other lighter mesons and hadrons. In the past quite
consciously, we never differentiated between initial state and final
state effects. Herein too we do the same. And finally now we will be
probing here whether there could be any special effect arising out
of a smaller system-size for $Cu+Cu$ compared to that for $Au+Au$
reactions.
\par
Our plan of work presented here is as follows. In the next section
we try to present a model not based on the concepts of quarks and
gluons; rather it is founded on altogether different facade
presented in details in the Refs\cite{bhat79}-\cite{bhat882}. In
section 3 we give the results and a general discussion on them. And
in the last section (section 4) we offer the final remarks which
essentially summarise the conclusions of this work.
\section{The Model and the Approach}
The description of the model-based features would be subdivided into
two parts. The first part gives a brief overview of the
$J/\Psi$-production mechanism in nucleon-nucleon ($p+p$) interaction
in the context of the Sequential Chain Model (SCM). Thereafter, the
relevant transition for different observables from $p+p$ to $A+B$
interactions will be discussed.
\par
The outline and characteristics of the model we use here are
obtained, in the main, from one of our previous works \cite{pgr05}.
According to this model, the photons released by two dominant
varieties of vector mesons which are produced in a
$\varrho$-$\omega$-$\pi$ sequential chain with $\varrho$ (rho) and
$\omega$ (omega) alternating with pion(s). According to this model,
called Sequential Chain Model (SCM),  high energy hadronic
interactions boil down, essentially, to the pion-pion interactions;
as the protons are conceived in this model as
$p$=($\pi^+$$\pi^0$$\vartheta$), where $\vartheta$ is a spectator
particle needed for the dynamical generation of quantum numbers of
the nucleons \cite{bhat79}-\cite{pgr072}. The particle $\vartheta$
is called a spectator particle because it does not and cannot take
part in the strong interactions, as it is identified to be a muonic
neutrino and is taken to be a Majorana spinor. The internal quantum
numbers like isospin, strangeness and baryon numbers can be related
to the internal angular momentum of the constituents and we thus do
have a geometrical origin of the basic SU(3) symmetry of the
hadrons. And it is taken that these constituents move in a harmonic
oscillator potential with orbital angular momentum $1/2\hbar$ in
such a way that, the two values of the third component of the
orbital momentum represent the two states of matter: particles and
antiparticles. The multiple production of $J/\Psi$-mesons in a high
energy proton-proton collisions is described in the following way.
The secondary $\pi$-meson or the exchanged $\varrho$-meson emit a
free $\omega$-meson and pi-meson; the pions so produced at high
energies could liberate another pair of free $\varrho$ and trapped
$\omega$-mesons (in the multiple production chain). These so-called
free $\varrho$ and $\omega$-mesons decay quite a fast into photons
and these photons decay into $\Psi$ or $\Psi'$ particles, which,
according to this alternative approach is a bound state of $\Omega
\bar{\Omega}$ or $\Omega' \bar{\Omega}'$ particles. The production
of $J/\Psi$-mesons is shown schematically in Figure 1. The field
theoretical calculations for the average multiplicity of the
$\Psi$-secondaries and for the inclusive cross-sections of the
psi-particles deliver some expressions which we would pick up from
\cite{pgr05} and \cite{sanyal}.
\par
The inclusive cross-section of the $\Psi$-meson produced in the $p+p$ collisions given by
\begin{equation}\displaystyle{
E\frac{d^3\sigma}{{dp}^3}|_{p+p\rightarrow{{J/\Psi}+X}}  \cong
C_{J/\Psi}\frac{1}{p_T^{N_R}}\exp(\frac{-5.35(p_T^2+m^2_{J/\Psi})}{<n_{J/\Psi}>^2_{p+p}(1-x)})
\exp(-1.923{<n_{J/\Psi}>_{p+p}}x),}
\end{equation}
where the expression for the average multiplicity for
$\Psi$-particles in $p+p$ scattering is
\begin{equation}\displaystyle{
<n_{J/\Psi}>_{p+p} ~~~ = ~~~ 4\times10^{-6}s^{1/4}.}
\end{equation}
In the above expression for eqn (1) the term $|C_{J/\Psi}|$ is a
normalisation parameter and is assumed here to have a value $\cong
0.09$ for Intersecting Storage Ring(ISR)energy, and it is different
for different energy region and for various collisions. The terms
$p_T$, $x$ and $m_{J/\Psi}$ represent the transverse momentum,
Feynman Scaling variable and the rest mass of the $J/\Psi$ particle
respectively. Moreover, by definition, $x ~ = ~ 2p_L/{\sqrt s}$
where $p_L$ is the longitudinal momentum of the particle. The $s$ in
equation (2) is the square of the c.m. energy.
\par
The second term in the right hand side of the equation (1), the
constituent rearrangement term arises out of the partonic
rearrangements inside the proton. But this proposed `rearrangement'
factor is different from what are implied by the terms `parton
recombination' or `parton coalescence' in the standard literatures
on $\Psi$-studies. In fact, this is also different from what was
meant by parton rearrangement by Bleibel et al. \cite{bleibel}.  It
is established that hadrons(baryons and mesons) are composed of few
partons. At large transverse momenta in the high energy interaction
processes the partons undergo some dissipation losses due to the
impact and impulse of the projectile on the target and the parton
inside them (both the projectile and the target) they suffer some
forced shifts of their placements or configurations. These
rearrangements mean undesirable loss of energy , in so far as the
production mechanism is concerned. The choice of ${N_R}$ would
depend on the following factors: (i) the specificities of the
interacting projectile and target, (ii) the particularities of the
secondaries emitted from a specific hadronic or nuclear interaction
and (iii) the magnitudes of the momentum transfers and of a phase
factor (with a maximum value of unity) in the rearrangement process
in any collision. And this is a factor for which we shall have to
parametrise alongwith some physics-based points indicated earlier.
The parametrisation is to be done for two physical points, viz., the
amount of momentum transfer and the contributions from a phase
factor arising out of the rearrangement of the constituent partons.
Collecting and combining all these, we proposed the relation to be
given by \cite{pgr072}
\begin{equation}\displaystyle
N_R=4<N_{part}>^{1/3}\theta,
\end{equation}
where $<N_{part}>$ denotes the average number of participating
nucleons and $\theta$ values are to be obtained phenomenologically
from the fits to the data-points. In this context, the only
additional physical information obtained from the observations made
here is: with increase in the peripherality of the collisions the
values of $\theta$ gradually grow less and less, and vise versa.
$\theta$ is a phase term related to the geometry and number of the
rearrangement of the partons of the nucleus. And the number of
rearrangements is constrained by the impact parameter which decides
the centrality/peripherality of the collisions reflected by
$N_{part}$.
\par
The complex calculations of the Feynman diagram with the infinite momentum frame techniques give rise to
such forms with some assumptions and approximations based on the boundedness of $p_T$ (transverse momentum)-values.
For calculation purposes we took $f_{\varrho\omega\pi}$, $g_{\gamma\omega}$ and $g_{\gamma\Psi}$ as the point-coupling
of Fig.1. Moreover, in this context, we would have to reckon some other model-dependant features: (1) There are total
nine probable diagrams for the various exchanges of $\pi\pi\rightarrow\pi\pi$. So the final contribution has a product
term of $9^2$; (2) Secondly, the c.m. energy of the $\pi\pi$ system has to be translated into the actual $p+p$
collisions for which one derives a conversion factor $s_{\pi\pi}={\frac{4}{25}}s_{pp}$. The previous paper \cite{bhat79}
does not contain the product term (i.e. the factor) $1/p_T^{N_R}$. This multiplier term has been introduced
from physical considerations arising out of large-$p_T$ (hard) reactions at RHIC energies. in fact, this term
has been inducted absolutely phenomenologically. But the rest is an outcome of very rigorous calculations based
on `soft' (small-$p_T$) production of particles. In nature 90$\%$ of the particles are produced with very small $p_T$.
And our initial focus during the period 1975-1980 was to concentrate on generalized particle-production characteristics
in nature.
\par
In order to study a nuclear interaction of the type
$A+B\rightarrow Q+ x$, where $A$ and $B$ are projectile and target
nucleus respectively, and $Q$ is the detected particle which, in the
present case, would be $J/\Psi$-mesons, the SCM
has been adapted, on the basis of the suggested Wong \cite{wong} work to the
Glauber techniques by using Wood-Saxon
distributions \cite{eskola}-\cite{gorenstein}. The inclusive
cross-sections for $J/\Psi$ production in different nuclear
interactions of the types $A+B\rightarrow J/\Psi+ X$ in the
light of this modified Sequential Chain Model (SCM) can then be written
in the following generalised form as \cite{pgr031}-\cite{pgr032}:
\begin{equation}\displaystyle
{E{\frac{d^3\sigma}{dp^3}}|_{AB\rightarrow{J/\Psi}+ X}=
P_{J/\Psi} {p_T}^{-N_R} \exp(-c(p_T^2+m^2_{J/\Psi}))
\exp(-1.923{<n_{J/\Psi}>_{pp}x)}}.
\end{equation}
 where $P_{J/\Psi}$, $N_R$ and $c$ are
the factors to be calculated under certain physical constraints.
With the details of the calculations to be obtained from
Refs.\cite{pgr031}-\cite{pgr032}, the set of relations to be used
for evaluating the parameters $P_{J/\Psi}$  is given below.
\begin{equation}\displaystyle{
P_{J/\Psi}=C_{J/\Psi}{\frac{3}{2\pi}}{\frac{(A \sigma_B + B
\sigma_A)}{\sigma_{AB}}}
{\frac{1}{1+\alpha(A^{1/3}+B^{1/3})}}}
\end{equation}
Here, in the above set of equations, the third factor gives a
measure of the number of wounded nucleons i.e. of the probable
number of participants, wherein $A\sigma_B$ gives the probability
cross-section of collision with `$B$' nucleus (target), had all
the nucleons of $A$ suffered collisions with $B$-target. And
$B\sigma_A$ has just the same physical meaning, with $A$ and $B$
replaced. Furthermore, $\sigma_A$ is the
nucleon(proton)-nucleus(A) interaction cross section, $\sigma_B$
is the inelastic nucleon(proton)-nucleus(B) reaction cross section
and $\sigma_{AB}$ is the inelastic $AB$ cross section for the
collision of nucleus $A$ and nucleus $B$. The values of
$\sigma_{AB}$, $\sigma_{A}$, $\sigma_{B}$ are worked here out in a
somewhat heuristic manner by the following formula \cite{na5002}
\begin{equation}
\displaystyle{ \sigma^{inel}_{AB} ~ = ~ \sigma_{0} ~
(A^{1/3}_{projectile} + A^{1/3}_{target} - \delta)^2}
\end{equation}
with $\sigma_{0} = 68.8$ mb, $\delta= 1.32$.
\par
Besides, in expression (5), the fourth term is a physical factor
related with energy degradation of the secondaries due to multiple
collision effects. The parameter $\alpha$ occurring in eqn.(5) above
is a measure of the fraction of the nucleons that suffer energy
loss. The maximum value of $\alpha$ is unity, while all the nucleons
suffer energy loss. This $\alpha$ parameter is usually to be chosen
\cite{wong}, depending on the centrality of the collisions and the
nature of the secondaries.
\par
The ``$P_{J/\Psi}$" factor in the expression (5) accommodates a wide
range of variation because of the existence of the large differences
in the in the normalizations of the $J/\Psi$ cross-sections for
different types of interactions. The constraints for calculating
$P_{J/\Psi}$ are mentioned above.
\par
The parameter `$N_R$' of Eqn. (4) has been calculated from the
Eqn.(3). The constraints for calculating $N_R$ have been stated in
the paragraph written below the Eqn. (4).
\par
The parameter `$c$' is a energy and rapidity dependent term. It is
clear from eqn. (1) that energy dependent factor comes from the
average multiplicity of $J/\Psi$ particles($\sim s^{-0.125}$)and the
rapidity factor comes from $(1-x)$ where $x= \frac{2m_T \sinh
y_{cm}}{\sqrt s}$. So, $c$ has different values for different
energies and rapidity regions.
\par
The parameters used in this work are (i)$C _{J/\Psi}$ is a
normalization term which is by nature species-dependent and
energy-dependent; (ii) $N_R=4<N_{part}>^{1/3}\theta$; this is
determined by fits to the invarient cross-section data. It is
naturally found to be both secondary-species and
centrality-dependent; (iii) Finally, $\alpha$ is the fraction of the
nucleus that suffer energy by undergoing multiple collisions and
this $\alpha$-factor too has weak centrality dependence.
\par
The parameters are generally obtained from the minimization of
$\chi^2$ of the experimental spectrum for invariant cross-section
which involves some rigorous statistical methods and complications.
But the main issues on parameter-constraining are the following few
points: (i) availability of the large number of experimental
data-points; (ii) smallness of the uncertainty-ranges in the
measured data. Unfortunately, for $J/\Psi$ production, in any high
energy collision whatsoever, the number of data points are
relatively sparse, especially for the large transverse momentum
values. Secondly, the uncertainty ranges in the measurements are
also quite considerable, for which we completely dropped the idea
and bid to devise the methods of constraining the evolved
parameters. Very recently, Arleo and d`Enterria \cite{arleo} studied
on constraining the parameters for very simple and abundantly
produced neutral pion production in $pp$-scattering at $\sqrt
s$=22.4 GeV and ended up with a statement like `` At high $p_T$'s
the fit is completely unconstrained due to the lack of data, and its
uncertainty is very large."
\section{Model-based Analyses and the Results}
Here, what we need to emphasize is that the observables are
different at the same energy. So let us split up the cases and
treat the data available on them, individually and on a case to
case basis.
\subsection{$J/\Psi$ (total) Crosssections in $p+p$ Interactions}
As the psi-productions are generically treated rightly as the
resonance particles, the standard practice is to express the
measured $J/\Psi$ (total) crosssections times branching ratio to
muon or electrons, i.e.for lepton pairs , that is by
$B_{ll'}\sigma^{J/\Psi}_{p+p}$ \cite{gonin}. And this is usually done for
all the symmetric nuclear collision like $Au+Au$, $Pb+Pb$ and
$Cu+Cu$ etc.
\par
By using expression (4) we arrive at the expressions for the
differential cross-sections for the production of $J/\Psi$-mesons
in the mid and forward-rapidities (i.e. $|y|<0.35$ and
$1.2<|y|<2.2$ respectively) in $p+p$ collisions at
$\sqrt{s_{NN}}$=200 GeV at RHIC.
\begin{equation}\displaystyle{
\frac{1}{2\pi p_T} B_{ll'} \frac{d^2\sigma}{dp_Tdy}|_{p+p\rightarrow J/\Psi+X} = 6.1 p_T^{-1.183} \exp[-0.13(p_T^2+9.61)]~~~~ for ~~|y|<0.35,}
\end{equation}
and
\begin{equation}\displaystyle{
\frac{1}{2\pi p_T} B_{ll'} \frac{d^2\sigma}{dp_Tdy}|_{p+p\rightarrow
J/\Psi+X} = 6.5 p_T^{-1.183} \exp[-0.16(p_T^2+9.61)]~~~~ for
~~1.2<|y|<2.2.}
\end{equation}
For deriving the expressions (7) and (8) we have used the relation
$x\simeq \frac{2p_{Zcm}}{\sqrt s}= \frac{2m_T \sinh y_{cm}}{\sqrt
s}$ \cite{pdg}, where $m_T$ , $y_{cm}$ are the transverse mass of
the produced particles and the rapidity distributions. $m_{J/\Psi}
\simeq 3096.9\pm 0.011 MeV$ \cite{pdg} and $B_{ll'}$, the branching
ratio is for muons or electrons i.e. its for lepton pairs $J/\Psi \rightarrow \mu^+\mu^-/e^+e^-$, is taken as
$5.93\pm 0.10\times10^{-2}$ \cite{pdg} in calculating
the above equations.
\par
In Figure (2), we have drawn the solid lines depicting the
model-based results with the help of above two equations (7) and
(8) against the experimental measurements \cite{adare},\cite{leitch}.
\subsection{Invariant yields of $J/\Psi$ Particles in $d+Au$, $Cu+Cu$ and $Au+Au$ Collisions}
From the expression (4), we arrive at the invariant yields for the $J/\Psi$-production in $d+Au\rightarrow J/\Psi+X$ reactions for mid and forward-rapidities.
\begin{equation}\displaystyle{
\frac{1}{2\pi p_T}  \frac{d^2N}{dp_Tdy}|_{d+Au\rightarrow J/\Psi+X} = 7.25\times10^{-7} p_T^{-0.629} \exp[-0.13(p_T^2+9.61)]~~~~ for ~~|y|<0.35,}
\end{equation}
and
\begin{equation}\displaystyle{
\frac{1}{2\pi p_T}  \frac{d^2N}{dp_Tdy}|_{d+Au\rightarrow J/\Psi+X} = 4.25\times10^{-7} p_T^{-0.629} \exp[-0.16(p_T^2+9.61)]~~~~ for ~~1.2<|y|<2.2.}
\end{equation}
In Figure (3), we have drawn the solid lines depicting the
model-based results with the help of above two equations (9) and
(10) against the experimental measurements \cite{adare4}.
\begin{table}
\begin{center}
\caption{Values of $b_{J/\Psi}$ and $N_R$ for different centrality
regions in $Cu+Cu$ collisions}
\begin{tabular}{ccccc}
\hline
Centrality [$|y|<0.35$]&0-20$\%$&20-40$\%$&40-60$\%$&60-92$\%$\\
\hline $b_{J/\Psi}$&$6.40\times10^{-5}$&$3.82\times10^{-5}$&$1.22\times10^{-5}$&$2.10\times10^{-6}$\\
$N_R$&0.829&0.709&0.625&0.615\\ \hline \hline
Centrality [$1.2<|y|<2.2$]&0-20$\%$&20-40$\%$&40-60$\%$&60-92$\%$\\
\hline $b_{J/\Psi}$&$6.38\times10^{-5}$&$2.82\times10^{-5}$&$0.82\times10^{-5}$&$1.90\times10^{-5}$\\
$N_R$&0.829&0.709&0.625&0.615\\
\hline
\end{tabular}
\end{center}
\end{table}
\par
For the case of $Cu+Cu$ collisions at RHIC, the values of
$b_{J/\Psi}=P_{J/\Psi}\times\exp(-1.923{<n_{J/\Psi}>_{pp}}x)${\footnote{The exponential part, here, contains $s$ and $x$ i.e.
the squared c.m. energy and the rapidity factor $y_{cm}$. This, in
course of calculations, gives some fixed values and is multiplied
with $P_{J/\Psi}$ to give the final product $b_{J/\Psi}$.} and
$N_R$ of expression (4) are given in Table 1 for the rapidities
$|y|<0.35$ and $1.2<|y|<2.2$ respectively and they are plotted in
Figure 3. The exponential parts for different rapidities are remain
the same as in equation (7) and equation (8). As previously stated,
the exponents $N_R$ of $p_T$ depend on the following three factors
mentioned in the previous section. The values of $<N_{part}>$  of
equation (3) for different centralities of $Cu+Cu$ collisions have
been taken from \cite{alver08} and the calculations have been done
accordingly. The experimental results for the invariant yields of
$J/\Psi$ production as a function of transverse momenta
\cite{adare2} at different centrality values ranging from $0-20\%$
to $60-92\%$ are plotted in Fig. (4).  The solid lines in the
figures show the model-based results.
\par
Similarly, for $Au+Au$ collisions at $\sqrt{s_{NN}}$=200 GeV at RHIC, the values of $b_{J/\Psi}$ and $N_R$  of equation (4) for different centrality regions are shown in Table 2 and the model-based results are plotted in Fig.5. For calculating the values of $N_R$, we have used the values of $<N_{part}>$ from \cite{adare3}. The
experimental results for the invariant yields of $J/\Psi$
production as a function of transverse momenta  at
different centrality values ranging from $0-20\%$ to $60-92\%$ and for
the rapidities $|y|<0.35$ and $1.2<|y|<2.2$ respectively are
taken from  the PHENIX Collaboration \cite{adare3}.
\bigskip
\begin{table}
\begin{center}
\caption{ Values of
$b_{J/\Psi}$ and
$N_R$ for
 different centrality regions in $Au+Au$ collisions}
\begin{tabular}{ccccc}
\hline
Centrality [$|y|<0.35$]&0-20$\%$&20-40$\%$&40-60$\%$&60-92$\%$\\
\hline $b_{J/\Psi}$&$1.32\times10^{-3}$&$0.86\times10^{-3}$&$1.65\times10^{-4}$&$0.25\times10^{-5}$\\
$N_R$&1.023&0.954&0.834&0.732\\ \hline \hline
Centrality [$1.2<|y|<2.2$]&0-20$\%$&20-40$\%$&40-60$\%$&60-92$\%$\\
\hline $b_{J/\Psi}$&$0.91\times10^{-3}$&$0.42\times10^{-3}$&$0.19\times10^{-3}$&$0.22\times10^{-4}$\\
$N_R$&1.023&0.954&0.834&0.732\\
\hline
\end{tabular}
\end{center}
\end{table}
\subsection{Rapidity Distributions for Different reactions at ${\sqrt{s_{NN}}} = 200 GeV$}
For the calculation of the rapidity distribution from the set of
equations (1), (2), (3) and (4) we can make use of a standard relation as
given below:
\begin{equation}\displaystyle{
\frac{dN}{dy}=\int \frac{1}{2\pi p_T}\frac{d^2N}{dp_Tdy}dp_T}
\end{equation}
The rapidity distributions for the $J/\Psi$-production has now been reduced to a simple
relation stated hereunder
\begin{equation}\displaystyle{
\frac{dN}{dy}={a_1}\exp(-0.23\sinh{y_{cm}}).}
\end{equation}
The normalization factor $a_1$ depends on the centrality of the collisions and
is obvious from the nature of the eqn.(1), eqn. (2), eqn.(3) eqn.(4) and eqn. (11).
\par
For $p+p$ and $d+Au$ collisions, the calculated rapidity distribution equations are
\begin{equation}\displaystyle{
\frac{dN}{dy}|_{p+p\rightarrow J/\Psi+X}=1.215\times10^{-6}\exp(-0.23\sinh{y_{cm}}),}
\end{equation}
and
\begin{equation}\displaystyle{
\frac{dN}{dy}|_{d+Au\rightarrow J/\Psi+X}=7.025\times10^{-6}\exp(-0.23\sinh{y_{cm}}),}
\end{equation}
In Fig. 6 and Fig. 7 we have plotted the rapidity distributions for
$J/\Psi$-production in $p+p$ and $d+Au$ collisions respectively.
Data in those figures are taken from Ref. \cite{adare4} and the
lines show the theoretical outputs.      \par To calculate the
rapidity distribution for $Au+Au$ collisions we are taking into
account the different centrality regions by making use of Table 2 in
addition to Eqs. (4) and (11). The values of $a_1$ are different for
different centrality regions for $Au+Au$ collisions and they are
given in the Table 3.
\begin{table}
\begin{center}
\caption{ Values of $a_1\times 10^{-6}$  for
 different centrality regions for $Au+Au$ collisions}
\end{center}
\begin{center}
\begin{tabular}{ccccc}
\hline
Centrality &0-20$\%$&20-40$\%$&40-60$\%$&60-92$\%$\\
\hline $a_1$&0.472&0.033&0.011&0.002\\
 \hline
\end{tabular}
\end{center}
\end{table}
\par
The solid lines in the Fig. 8 depict the theoretical plots of $dN/dy$ vs. $y$ for
different centralities while the data for $Au+Au$ collisions are taken from Ref.\cite{adare2}.
\subsection{$<p_T>$ and $<p_T^2>$ Values}
Next we attempt at deriving model-based expressions for both $<p_T>$
and $<p_T^2>$. Of these twin observables the $<p_T>$ would be used
by us in obtaining the ratios $R_{AA}$-s for different participant
numbers, $N_{part}$. Though, very strangely, there is so far no data
on $<p_T>^{J/\Psi}$ even for $p+p$ collisions at ISR to RHIC, we
have chosen here to calculate and plot the nature of
$<p_T>^{J/\Psi}$ vs. $N_{part}$ for some collisions under study here
for the following reasons: (i) in our opinion, $<p_T>^{J/\Psi}$ is
more fundamental observable than $<p_T>^2$, (ii) for cases of almost
all the secondaries produced in high energy collisions it is found
that $<p_T>\neq<p_T^2>$ or $<p_T>^2\neq<p_T^2>$ and (iii) we find in
our work that $J/\Psi$ behaves quite similarly as many other
hadronic secondaries in nearly all the high energy interactions. So,
plots of $<p_T>$ vs. $N_{part}$ for $Cu+Cu$  and $Au+Au$ collisions
are shown in Fig. 9 mostly in a predictive vein with the hope that
it would be measured in future. And the excessive importance given
to $<p_T^2>^{J/\Psi}$ here by both the theorists and the
experimentalists is just contextual. But, from physical
considerations, one must admit, $<p_T>$ is much more important than
$<p_T^2>$. Furthermore, we also uphold the view that the theorists
should not be guided by just the situational constructs of the
experimentalists.
\par
The definition for average transverse momentum  $<p_T>^{J/\Psi}$ is
given below.
\begin{equation}\displaystyle{
<p_T>^Q=\frac{\int^{p_T(max)}_{p_T(min)}p_TE{\frac{d^3\sigma}{dp^3}}^Qdp_T^2}{\int^{p_T(max)}_{p_T(min)}E{\frac{d^3\sigma}{dp^3}}^Q
dp_T^2}, }
\end{equation}
The theoretical plot of the average transverse momentum
$<p_T>^{J/\Psi}$ in $Cu+Cu$ and $Au+Au$ collisions at RHIC for different number
of participating nucleons is shown in the Fig.9. The theoretical
calculations are done on the basis of uses of eqn.(4), Table 1 and
Table 2.
\par
The QCD-oriented models believe that the transverse momentum
distributions of the produced $J/\Psi$s carry valuable information
about the mechanism for their absorption or disappearance manifested
in the clearly falling nature of the production cross-section.
The ideas based on the Standard Model (SM) predict that there should
be a decrease of $<p_T^2>$ with centrality for sufficiently central
collisions which is viewed as a consequence of the colour
deconfinement `turnover effect'\cite{kharzeev1}, \cite{kharzeev2}.
The present model-dependent behaviour of the observables $<p_T^2>$ as is
defined by the undernoted relation (eqn. 16 given below). The
[$<p_T^2>$] is measured by RHIC and other experiments which is
supposed to have a close relationship with what-is-called-to-be the
`` $J/\Psi$ suppression", according to the QCD-points of view and to
some QCD-oriented models.
\par
 The expression for the average of the squared transverse momenta
for any secondary ($Q$), by definition, is
\begin{equation}\displaystyle{
<p_T^2>^Q=\frac{\int^{p_T(max)}_{p_T(min)}p_T^2E{\frac{d^3\sigma}{dp^3}}^Qdp_T^2}{\int^{p_T(max)}_{p_T(min)}E{\frac{d^3\sigma}{dp^3}}^Q
dp_T^2}, }
\end{equation}
Using the above definition and putting the form of inclusive
cross-section given by eqn.(4) into use, we calculate the
$<p_T^2>$ for the $J/\Psi$ production in $d+Au$, $Cu+Cu$ and $Au+Au$ collisions
collisions at RHIC energy $\sqrt{s_{NN}}$ = 200 GeV . The values of
$p_T(min)$ and $p_T(max)$ are taken here 0 and 5 respectively
\cite{adare4}, \cite{adare2}, \cite{adare3}.
\par
We are interested not only in the absolute values of $<p_T^2>$
alone, but also in the centrality-dependence of them. Because, the
conventional view holds the idea that $<p_T^2>$ has a specific form
of centrality-dependence, for which we present the graphical
representations of $N_{part}$ dependence of $<p_T^2>$ for the
rapidities $|y|<0.35$ and $1.2<|y|<2.2$ and for $d+Au$, $Cu+Cu$ and
$Au+Au$ collisions in Figures 10 (a) and (b) respectively. The
experimental values are taken respectively from \cite{adare4},
\cite{adare2}, \cite{adare3}.
\par
Moreover, in Table 4, we present $<p_T^2>$  calculated from the two
theoretical models, the SCM and  the `Hydro+$J/\Psi$' \cite{akc2},
with the experimental values \cite{adare2} for $p+p$ and $Au+Au$
collisions at RHIC energies.
\begin{table}
\begin{center}
\caption{ Comparison of $<p_T^2>$ values:: SCM vs. Hydro+$J/\Psi$
\cite{akc2} for different centrality bins and in mid-rapidity region
in  $p+p$ and $Au+Au$ collisions against the background of
experimental values.}
\end{center}
\begin{center}
\begin{tabular}{cccccc}
\hline
Reactions& Centrality&$N_{part}$&$<p_T^2>$ ($[GeV/c]^2$)&$<p_T^2>$ ($[GeV/c]^2$)&$<p_T^2>$ ($[GeV/c]^2$) \\
&&&(Expt.)&(SCM)&(Hydro+$J/\Psi$)\\
\hline $p+p$&&2&$4.1\pm0.2\pm0.1$&3.65&3.87\\
$Au+Au$&$0-20\%$&280&$3.6\pm0.6\pm0.1$& 3.64&3.76\\
&$20-40\%$&140&$4.6\pm0.5\pm0.1$&3.66&3.80\\
&$40-60\%$&60&$4.5\pm0.7\pm0.2$&3.55&3.81\\
&$60-92\%$&14&$3.6\pm0.9\pm0.2$&3.43&3.81\\ \hline
\end{tabular}
\end{center}
\end{table}
\subsection{The Nuclear Modifiacation Factor $R_{AA}$}
There is yet another very important observable called nuclear
modification factor (NMF), denoted here by $R_{AA}$ which for the
production of $J/\Psi$ is defined by \cite{adare2}
\begin{equation}\displaystyle{
R_{AA}=\frac{d^2N_{J/\Psi}^{AA}/dp_Tdy}{<N_{coll}(b)>d^2N_{J/\Psi}^{pp}/dp_Tdy }.}
\end{equation}
the SCM-based results  on NMFs for $Cu+Cu$ and $Au+Au$ collisions
are deduced on the basis of Eqn.(4), Eqn.(7), Eqn.(8) and Table 1
,Table 2 and they are  given by the undernoted relations
\begin{equation}\displaystyle{
R_{AA}|_{Cu+Cu\rightarrow J/\Psi+X}=0.42p_T^{0.35}~~~~ for ~~|y|<0.35,}
\end{equation}
\begin{equation}\displaystyle{
R_{AA}|_{Cu+Cu\rightarrow J/\Psi+X}=0.38p_T^{0.35},~~~~ for ~~1.2<|y|<2.2.}
\end{equation}
and
\begin{equation}\displaystyle{
R_{AA}|_{Au+Au\rightarrow J/\Psi+X}=0.36p_T^{0.16}~~~~ for ~~|y|<0.35,}
\end{equation}
\begin{equation}\displaystyle{
R_{AA}|_{Au+Au\rightarrow J/\Psi+X}=0.26p_T^{0.16},~~~~ for ~~1.2<|y|<2.2.}
\end{equation}
wherein the value of $<N_{coll}(b)>$  to be used is $\approx
170.5\pm 11$  \cite{alver05} for $Cu+Cu$ collisions and for $Au+Au$
collisions it is taken as $\approx 955.4\pm 93.6$  \cite{adler042}.
In Fig. 11, we plot $R_{AA}$ vs. $p_T$ for $0-20\%$ central region
in $Cu+Cu$ and $Au+Au$ collisions. The solid lines in the figure
show the SCM-based results against the experimental results
\cite{adare2}, \cite{adare3}.
\par
Next, for the analysis of centrality dependence as represented by
$(N_{part})$, of the nuclear modification factor, ($R_{AA}$) for
$Cu+Cu$  and $Au+Au$ collisions at RHIC, we proceed in the direction
with the help of eqn.(3), (4) and the calculated $<p_T>$s [Fig.5]
for forward rapidity and different centralities.  For $Cu+Cu$
collisions, the data in Figure 12 are taken from \cite{adare2}. The
solid line in that figure gives the SCM-based results. The dotted
lines in Figure 12 show the hydrodynamical (`Hydro+$J/\Psi$')
model-based calculations with and without nuclear absorption
(`hydro+j/Psi 1' and `hydro+J/Psi 2' respectively) \cite{akc}. And
for the $Au+Au$ collisions, the data in Figures 13(a) and (b) are
taken from \cite{adare3},\cite{zhao}. The solid line in that figure
gives the SCM-based results against the background of some other
predictions or calculations \cite{zhao}, \cite{vitev} \cite{akc}.
\subsection{LHC Prediction}
On the basis of the model-based calculations shown above, we now
make the prediction for the invariant yields of $J/\Psi$-mesons in
the most central (0-10$\%$)  $Pb+Pb$-collisions at the
$\sqrt{s_{NN}}=5500$ GeV i.e. at Large Hadron Collier (LHC) energy.
The equation of the invariant yields of $J/\Psi$ in the reaction of
the type $Pb+Pb\rightarrow J/\Psi+X$ for the forward rapidity region
would be of the form as given below
\begin{equation}\displaystyle{
\frac{1}{2\pi p_T} B_{ll'} \frac{d^2N}{dp_Tdy}|_{Pb+Pb\rightarrow J/\Psi+X} = 0.53\times10^{-2} p_T^{-1.123} \exp[-0.069(p_T^2+9.61)]~~ for ~~1.2<|y|<2.2.}
\end{equation}
In Figure 14, we had presented the predicted nature of the yields
for $J/\Psi$ production in $Pb+Pb$ collisions at LHC-energy by the
solid curve. The dashed plot in the figure represents our RHIC
calculations based on eqn.(4) and Table 2 in sect. 3.2.
\section{Summary and Conclusions}
In assessing the results achieved we must remember the limitations
of the circumstances and the issues. It has to be appreciated that
psions ($J/\Psi$-mesons) are not as abundant as pions, the commonest
variety of the secondaries. the cross-sections for
$J/\Psi$-production fall a few orders of magnitude compared to the
pions. Besides, the uncertainties in the measurements of these
psi-particles (resonance) are also much larger. So, in seeking the
agreements of the model-based results with the measured data on
$J/\Psi$-particles, one must be realistic. No model could probably
claim a cent-percent agreement with data under such circumstances as
pointed out above. For this reason, we have frequently compared here
our model-based results with both the experimental data and the
results obtained by one or two other competing models. There is yet
another point very specific to $J/\Psi$-particles. The number of
available and reliable data on psi-production in most cases is too
limited. So, we will have to cut down our expectations about the
degree of convergence between the calculations and the data and to
arrive at the justified comments on any model-based work on
$J/\Psi$. On this basis, we point out the following set of
observations on the present work. (i) The agreement between our
model-based (SCM-based) calculations and the $p_T$-dependence of the
invariant cross-sections on production $J/\Psi$-particles is
modestly satisfactory, for mid and forward rapidities,
moderate-$p_T$ and central nuclear collisions, though there are some
degrees of discrepancies for the other regions. This statement is
valid on an overall basis for the figures from Fig.2 -- Fig.5. (ii)
The plots on studies of rapidity dependence are somewhat constrained
by the very sparse data sets, for which we could make no strong or
emphatic comment about the nature of agreements or disagreements.
And this is a comment valid for the figures presented in Fig. 6 to
Fig. 8. (iii) We discussed in the text given above somewhat
extensively on the implications of the $<p_T^2>^{J/\Psi}$ and
$<p_T>^{J/\Psi}$-plots; our results, based on the Figs. (9) and (10)
and indicated by the Table-4 show fairly good agreement for
$<p_T^2>^{J/\Psi}$-data obtained by nuclear collisions and also with
`Hydro+$J/\Psi$' model \cite{akc2}. (iv) And the results shown by
Fig.(11)- Fig.(13) present a comprehensive study on the measured
observable $R_{AA}$ and the model-based results, including the
SCM-based plots in solid lines. (v) The Fig. 14 depicts our
SCM-based predictions for the behaviour of the observable at LHC
energy and in the LHC experiments at CERN.
\par
Thus, finally, we end up with a few conclusive remarks about the
physical significance of the work: (i) As the standard model does
not appear to be too promising for $J/\Psi$-production behaviour, we
have reasonably attempted a new one, surely with a certain degree of
tentativeness. (ii) Neither the measurements for $Cu+Cu$
interactions, nor the findings for $Au+Au$ reactions at RHIC
energies provide any final affirmation or clear indication of the
QGP Physics. This is what is established by the very recent studies
of both Tanenbaum \cite{tannenbaum},and Bhalerao, Gavai
\cite{bhalerao}. (iv) The underlying mechanism that has been used
here is not based on any `gluonisation'-concepts; so this approach
remains value-neutral to such theoretical inductions. (v) Nor does
this mechanism view the $J/\Psi$ particles as the products of the
heavy quarks, for which the ideas of `heavy quarks' cannot be
accommodated in this model; Similarly the phrases like `quark-gluon
plasma' (QGP), `anomalous suppression' etc. find no place in our
work. (vi) But, the new mechanism for $J/\Psi$-production suggested
by us, here incorporates the features of `normal' suppression by the
inclusion of the physical term of `partonic rearrangement factor'
yielding the normal suppression effect arising out mainly only at
large transverse momentum values belonging to `hard'-$p_T$
reactions. (vii) And for the `hard' sector of hadronic collisions
this model takes, without any exception, into account this
`constituent (or partonic) rearrangement factor' for production of
all the secondaries (light or heavy)in a consistent manner, though
certainly not on a uniform rate, which is certainly
secondary-specific in nature. Thus, the `globality property' of
particle production at high energy (and at large $p_T$) is upheld
for the production of even $J/\Psi$-mesons and so is not to be
treated on a different basis in our approach. (viii) The anomalous
observations \cite{prorok} on the centrality dependence of the
rapidity distributions very recently taken serious note of are kept
consciously out of the purview of the present work and would be
dealt with in a future work.
\newpage
\singlespacing

\newpage
\begin{figure}
\centering
\includegraphics[width=4in]{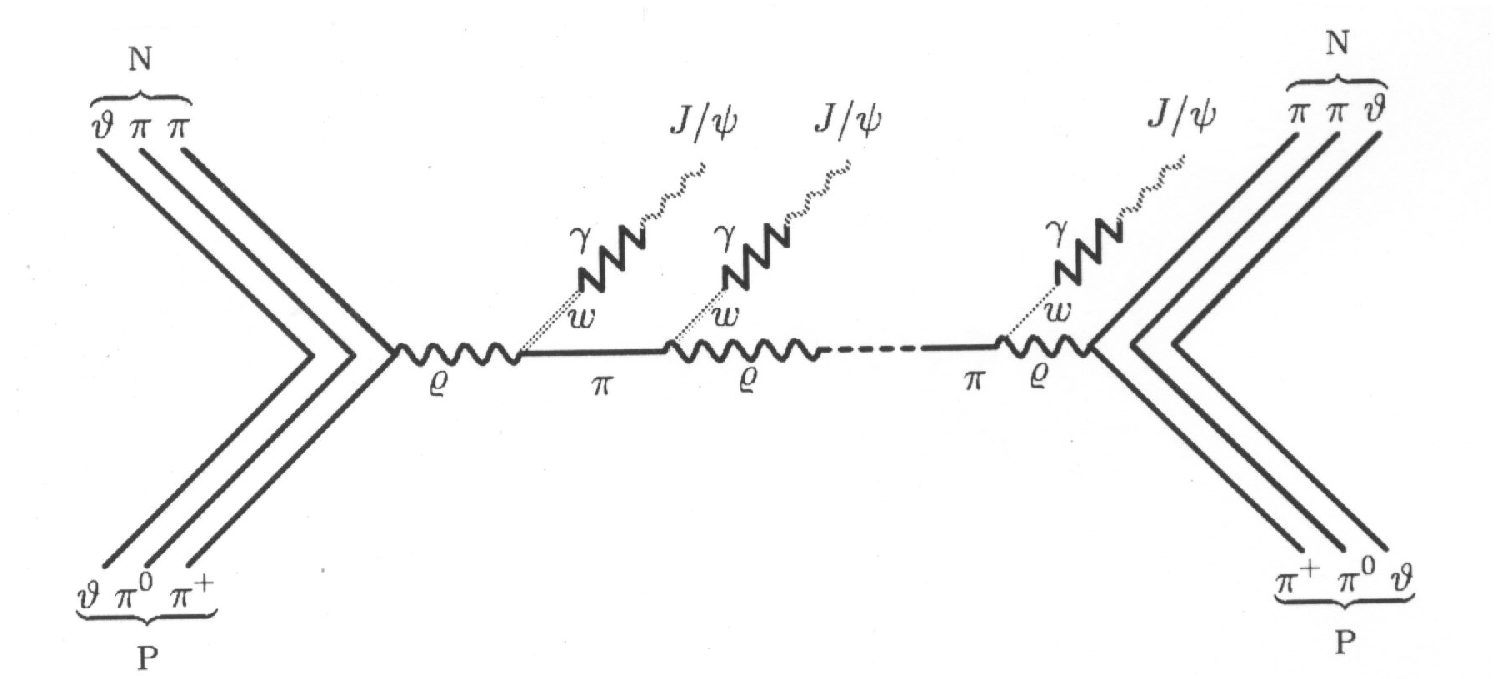}
 \caption{\small Schematic diagram of multiple production of $J/\Psi$ particles in $p+p$ scattering in the Sequential Chain Model.}
\end{figure}
\begin{figure}
\centering
\begin{minipage}{.5\textwidth}
\includegraphics[width=2.5in]{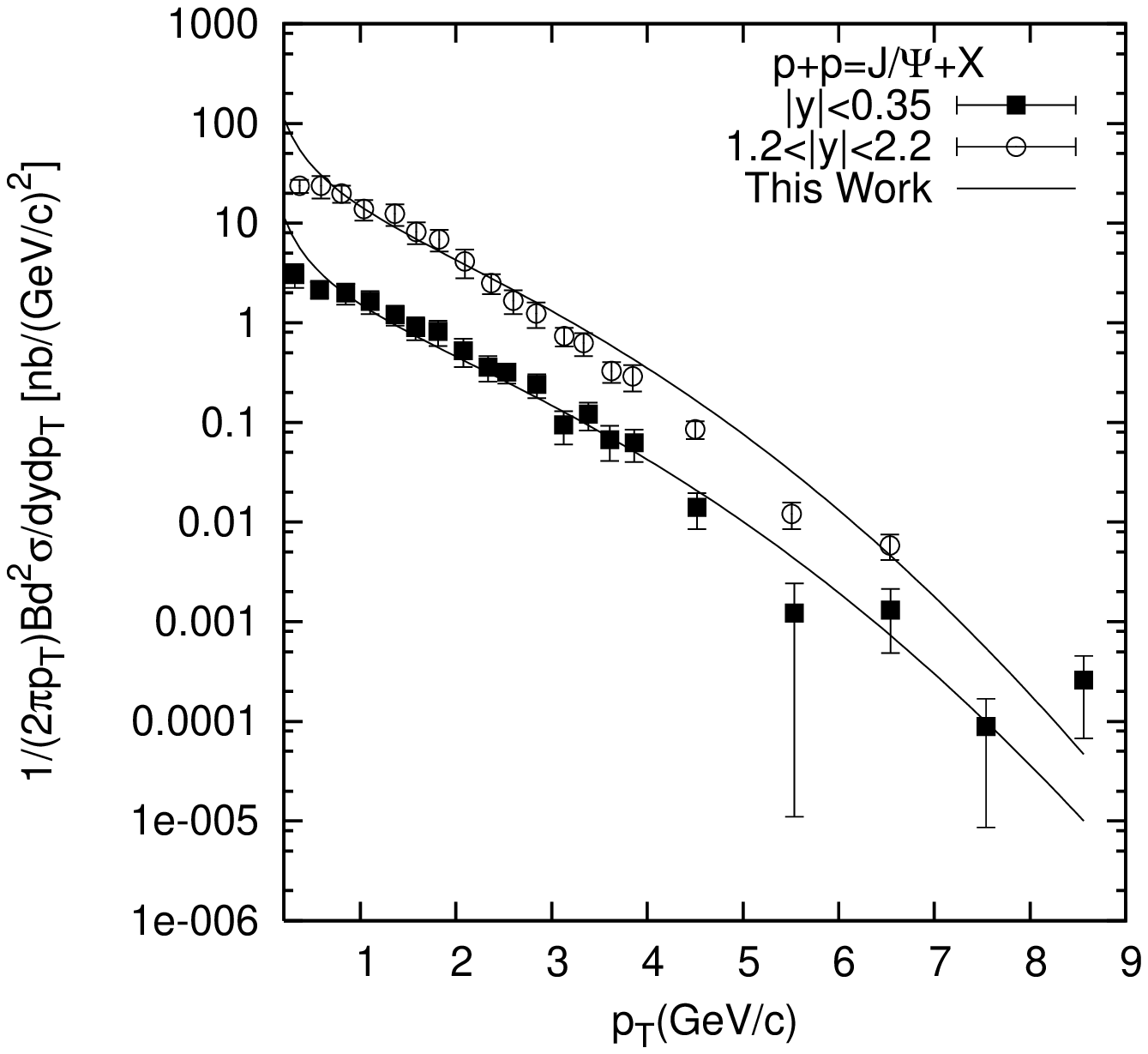}
\setcaptionwidth{3in} \caption{\small Plot of the invariant
cross-section for $J/\Psi$ production in proton-proton
collisions at $\sqrt s_{NN} =200 GeV$ as function of $p_T$. The
data points are from \cite{adare}. The solid curves
show the SCM-based results.}
\end{minipage}%
\begin{minipage}{.5\textwidth}
\centering
\includegraphics[width=2.5in]{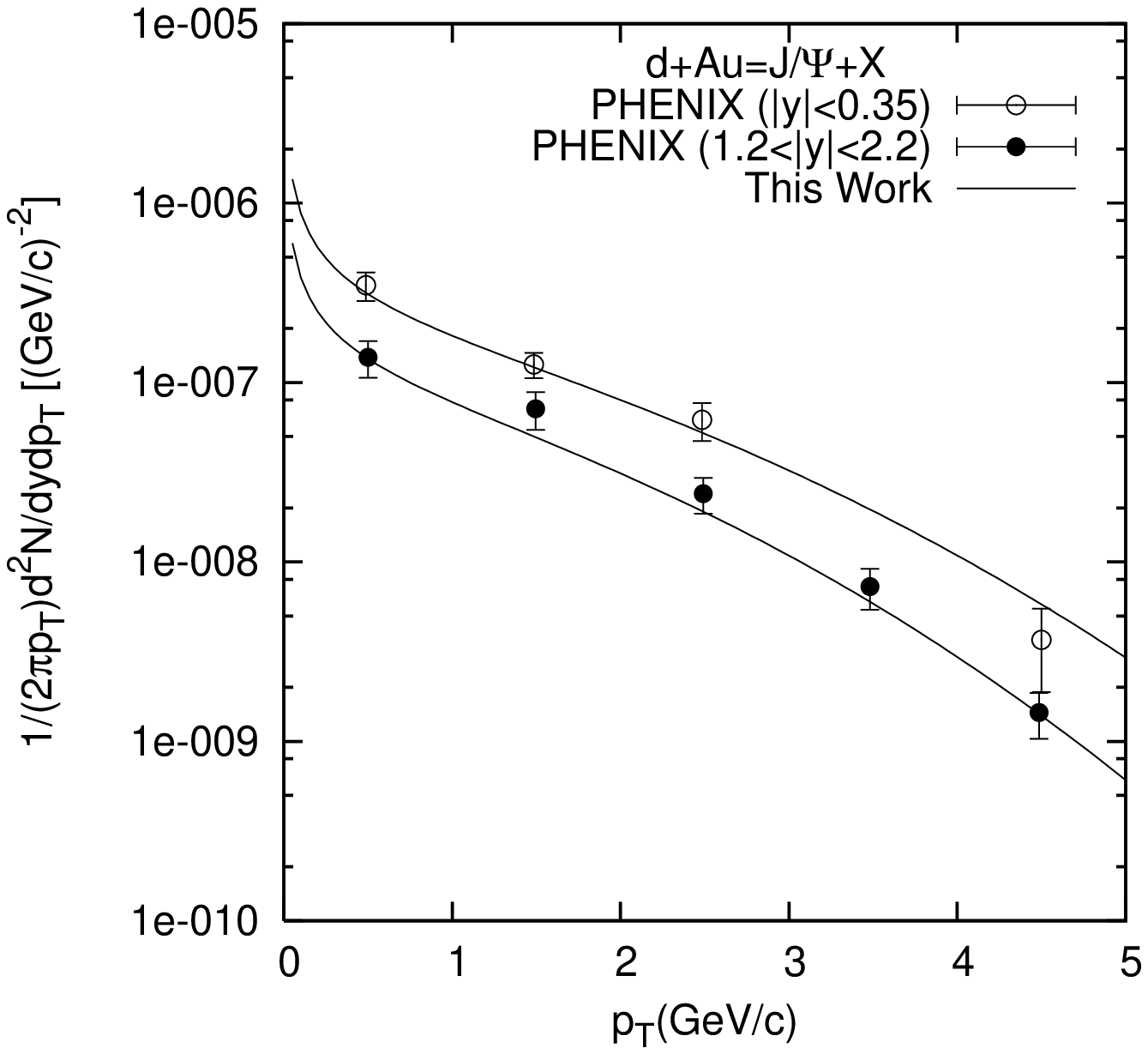}
\setcaptionwidth{3in} \caption{\small Plot of the invariant
cross-section for $J/\Psi$ production in $d+Au$ collisions at $\sqrt
s_{NN} =200 GeV$ as function of $p_T$. The data points are from
\cite{adare4}. The solid curves show the SCM-based results.}
\end{minipage}
\end{figure}
\begin{figure}
\subfigure[]{
\begin{minipage}{.5\textwidth}
\centering
\includegraphics[width=2.5in]{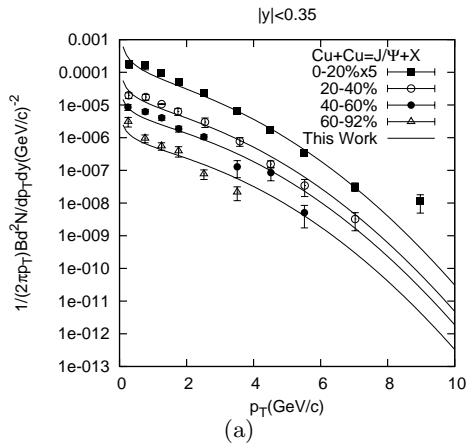}
\setcaptionwidth{2.6in}
\end{minipage}}%
\subfigure[]{
\begin{minipage}{0.5\textwidth}
\centering
 \includegraphics[width=2.5in]{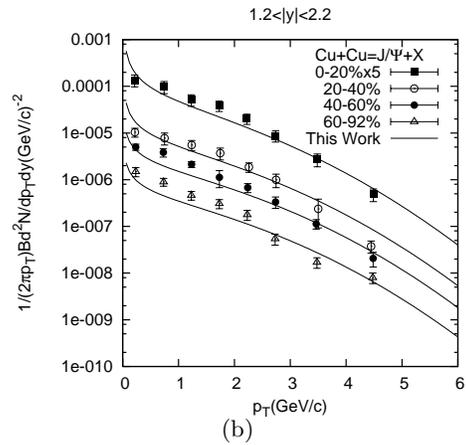}
 \end{minipage}}%
\caption{\small Transverse momenta spectra
at (a) $|y|<0.35$ and (b) $1.2<|y|<2.2$ for $J/\Psi$ production in $Cu+Cu$ collisions at
$\sqrt s_{NN} =200 GeV$. The
data are taken from \cite{adare2}. The solid curves
depict the SCM-based results.}
\end{figure}
\begin{figure}
\subfigure[]{
\begin{minipage}{.5\textwidth}
\centering
\includegraphics[width=2.5in]{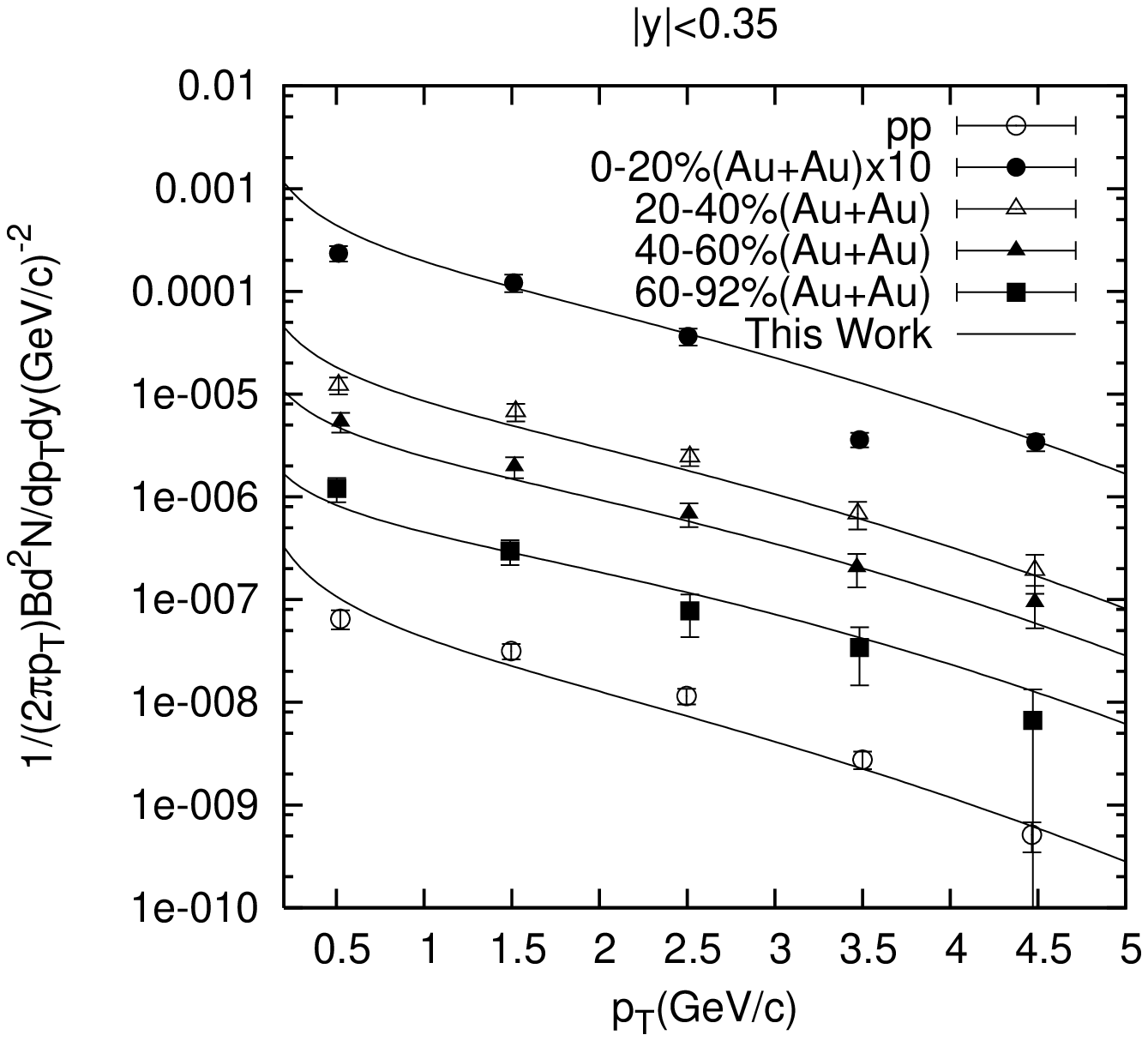}
\setcaptionwidth{2.6in}
\end{minipage}}%
\subfigure[]{
\begin{minipage}{0.5\textwidth}
\centering
 \includegraphics[width=2.5in]{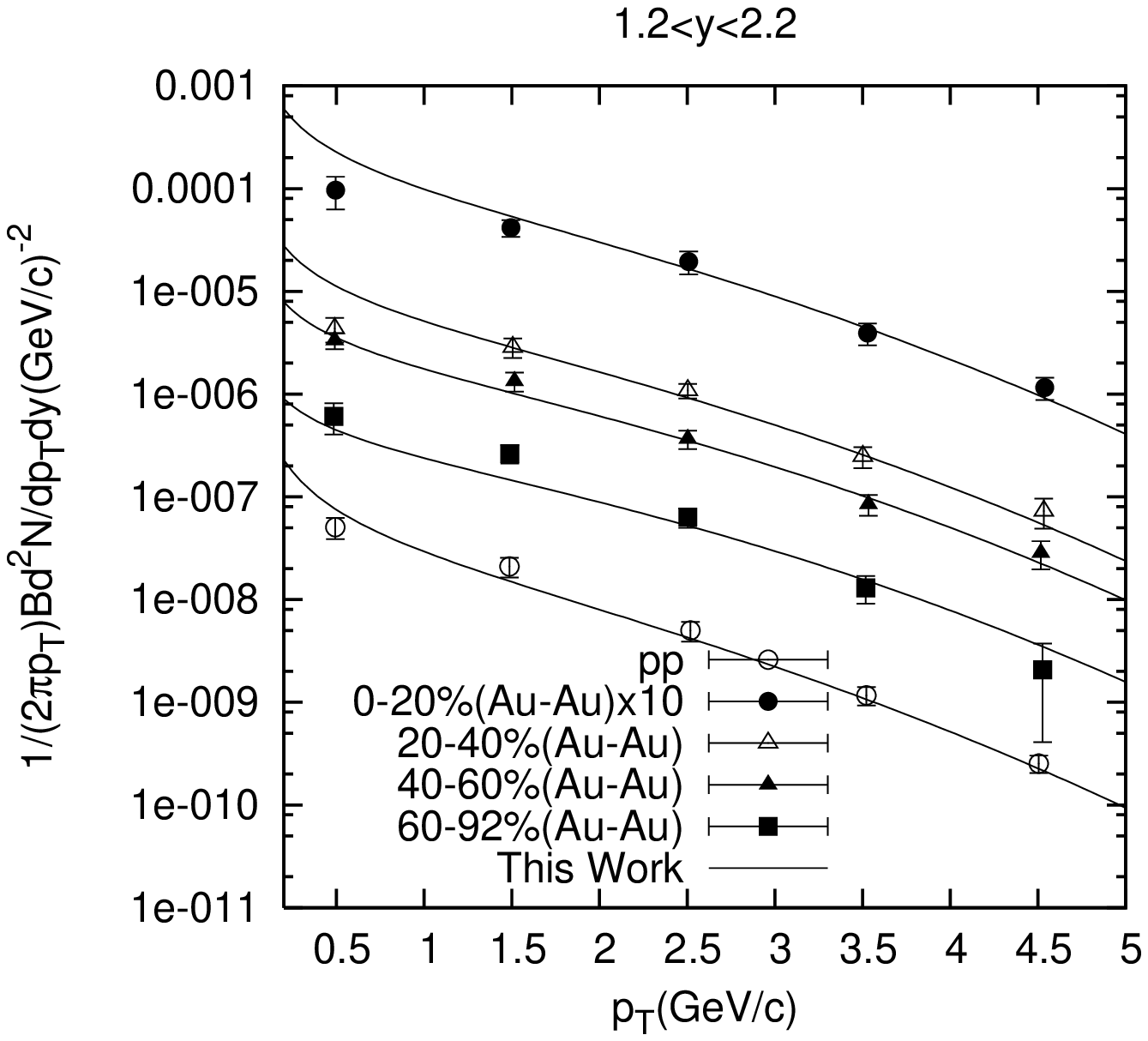}
 \end{minipage}}%
\caption{{\small Invariant
spectra as function of $p_T$ for $J/\Psi$ production in $Au+Au$
collisions at $\sqrt s_{NN} =200 GeV$  for (a) $|y|<0.35$ and (b) $1.2<|y|<2.2$ . The
data are taken from \cite{adare3}. The solid lines
show the SCM-based results.}  }
\end{figure}
\begin{figure}
\centering
\includegraphics[width=2.5in]{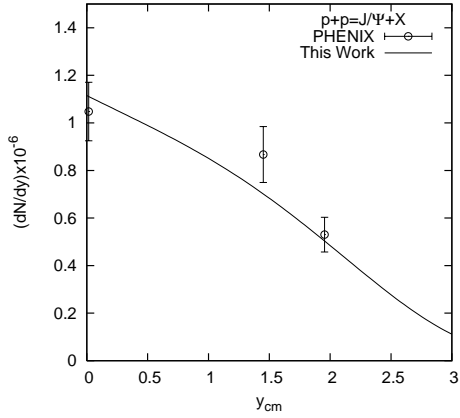}
 \caption{\small Plot of the rapidity distribution for $J/\Psi$ production in proton-proton
collisions at $\sqrt s_{NN} =200 GeV$ as function of $y$. The
data points are from \cite{adare4}. The solid curves
show the SCM-based results.}
\end{figure}
\begin{figure}
\centering
\includegraphics[width=2.5in]{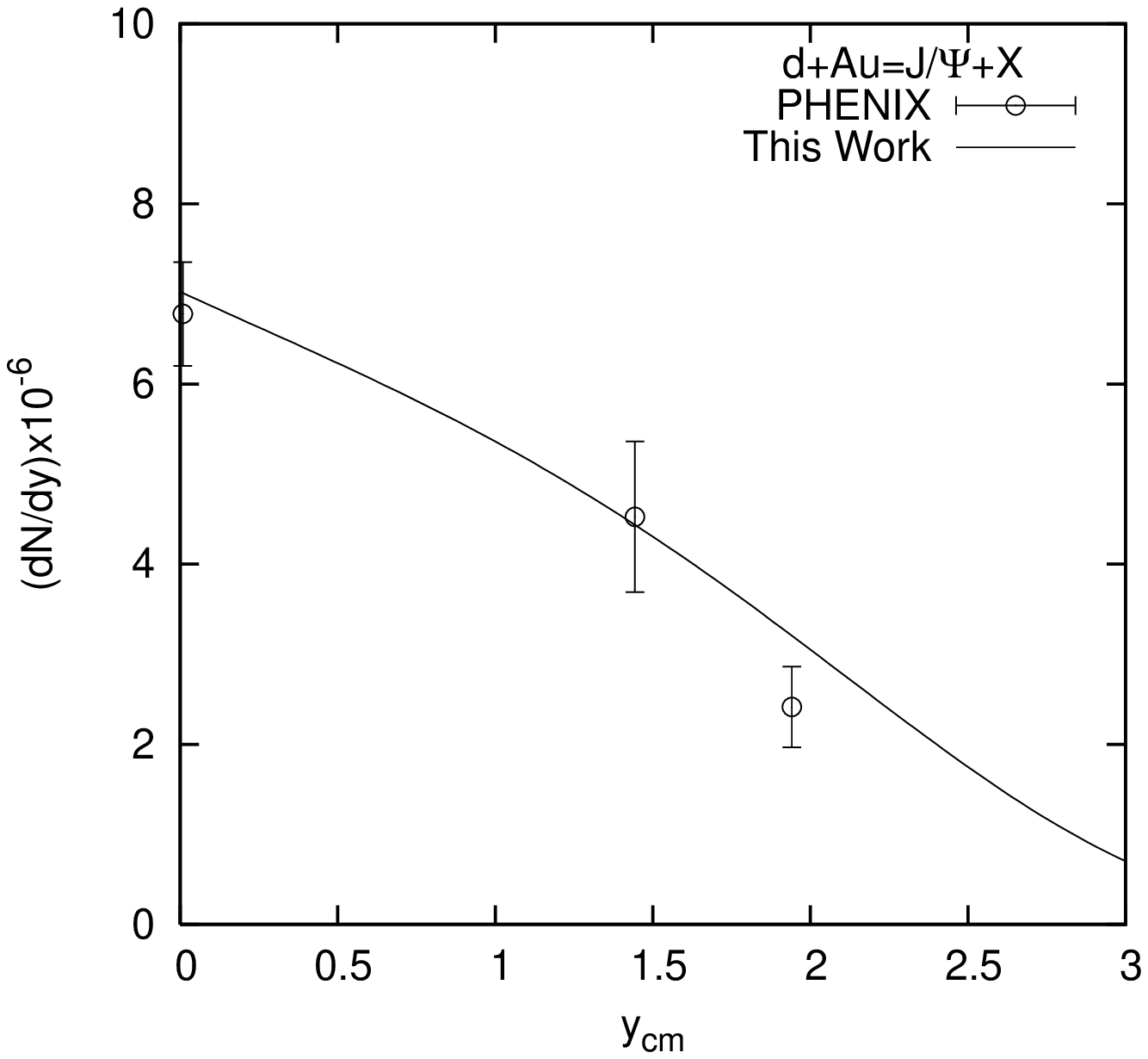}
 \caption{\small Rapidity distribution for
$J/\Psi$ production in $d+Au$ collisions at
$\sqrt s_{NN} =200 GeV$. The
data points are from \cite{adare4}. The solid curves
show the SCM-based results.}
\end{figure}
\begin{figure}
\subfigure[]{
\begin{minipage}{.5\textwidth}
\centering
\includegraphics[width=2.5in]{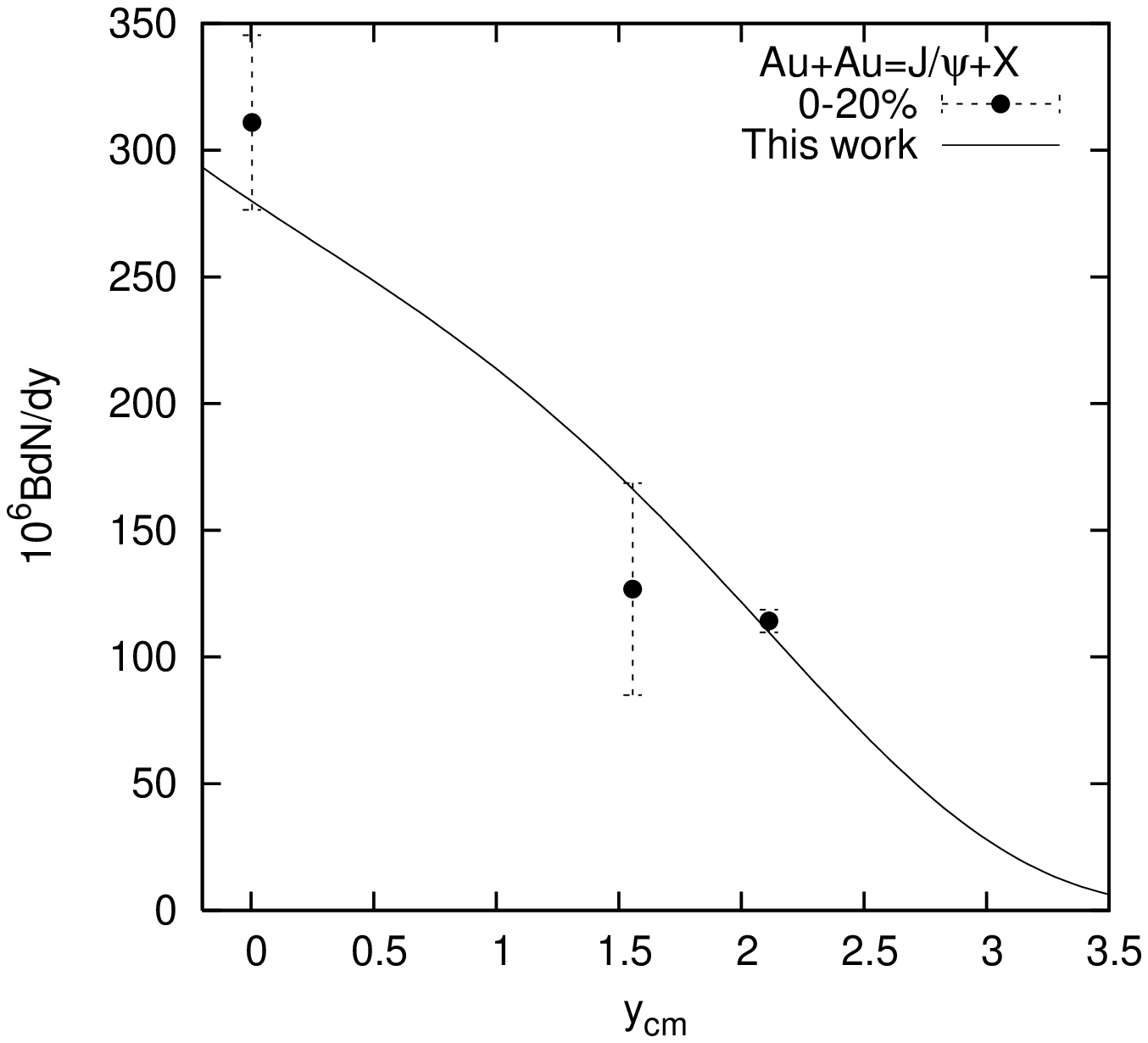}
\setcaptionwidth{2.6in}
\end{minipage}}%
\subfigure[]{
\begin{minipage}{0.5\textwidth}
\centering
 \includegraphics[width=2.5in]{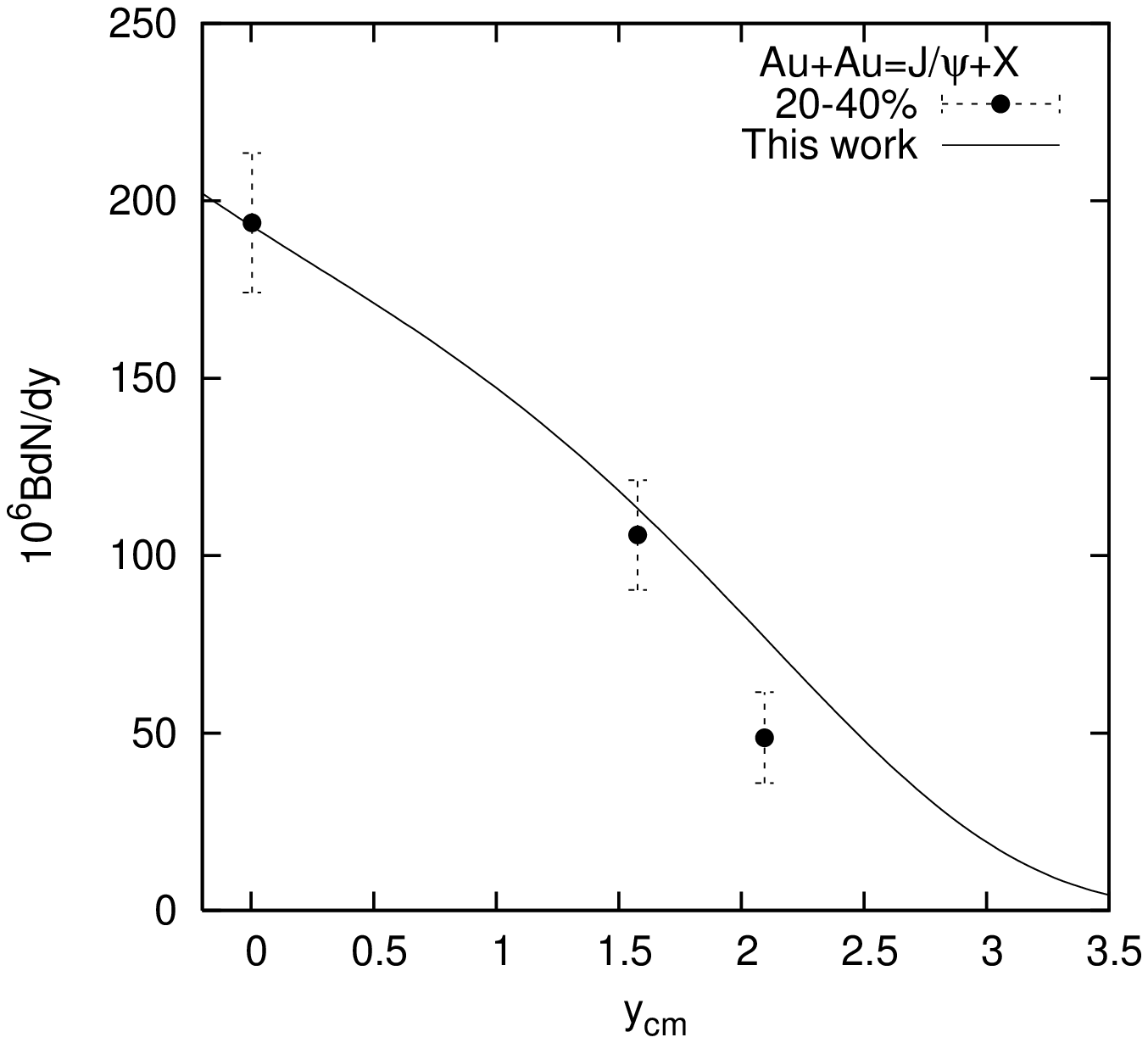}
  \end{minipage}}%
  \vspace{0.01in}
 \subfigure[]{
\begin{minipage}{0.5\textwidth}
\centering
 \includegraphics[width=2.5in]{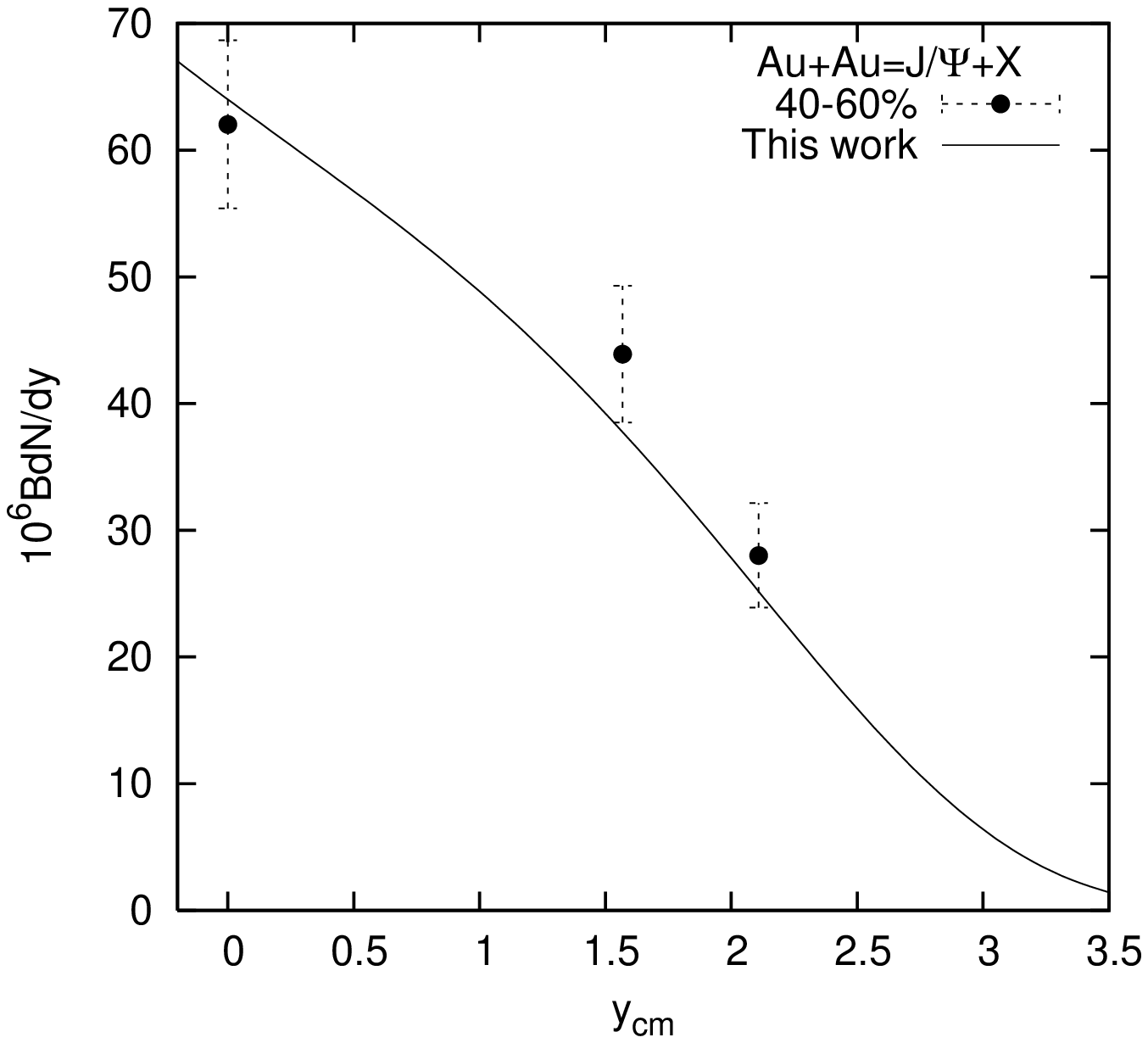}
 \end{minipage}}%
 \subfigure[]{
\begin{minipage}{0.5\textwidth}
\centering
 \includegraphics[width=2.5in]{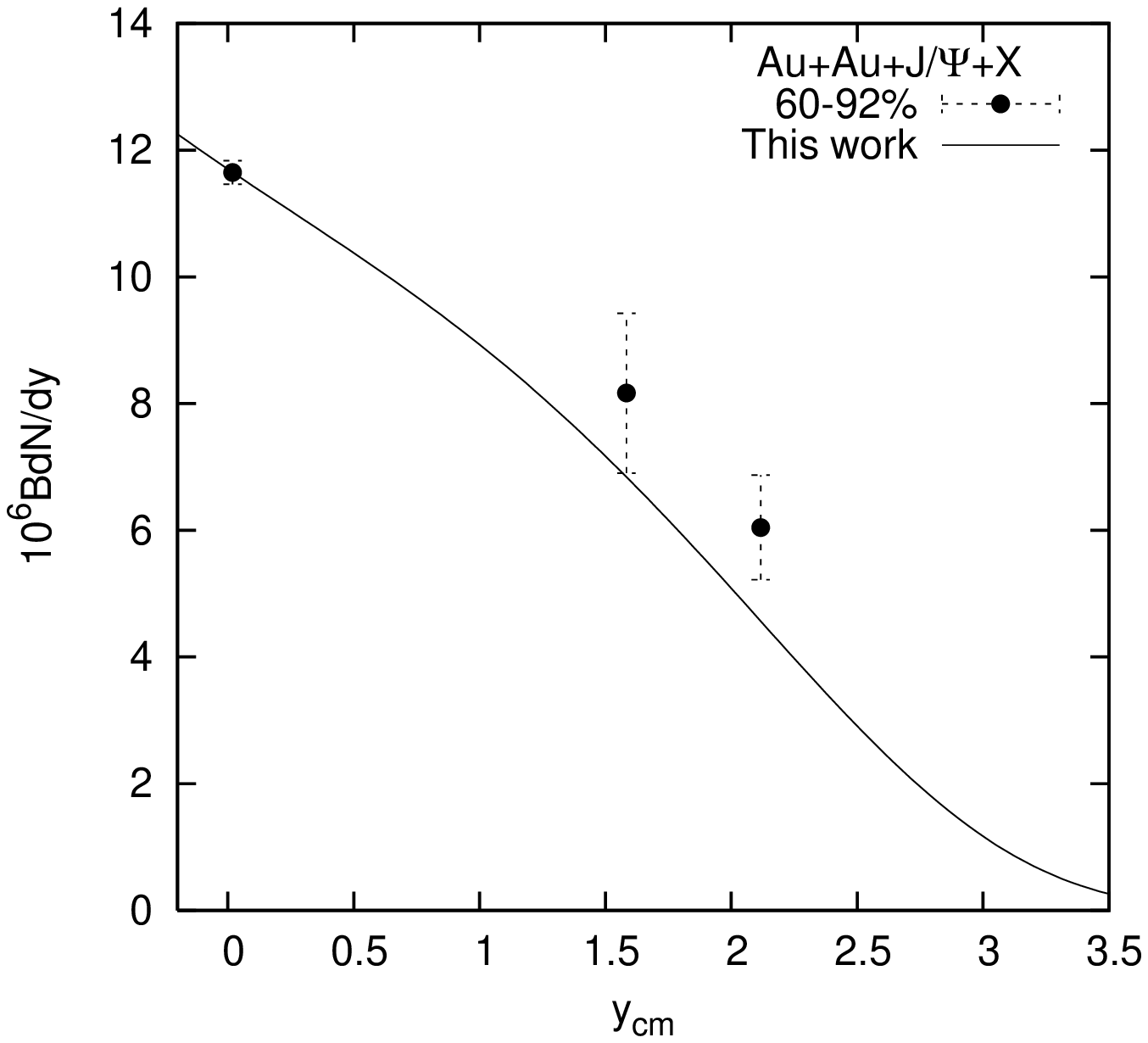}
 \end{minipage}}%
\caption{{\small Plot of $dn/dy$ vs. $y_{cm}$-values for  $Au+Au$
collisions at $\sqrt s_{NN} =200 GeV$ for the centrality widths
a)$0-20\%$, b)$20-40\%$, c)$40-60\%$ and d) $60-92\%$. The data
points are from \cite{adare2}. The solid line shows the SCM-based
results.}  }
\end{figure}
\begin{figure}
\subfigure[]{
\begin{minipage}{.5\textwidth}
\centering
\includegraphics[width=2.5in]{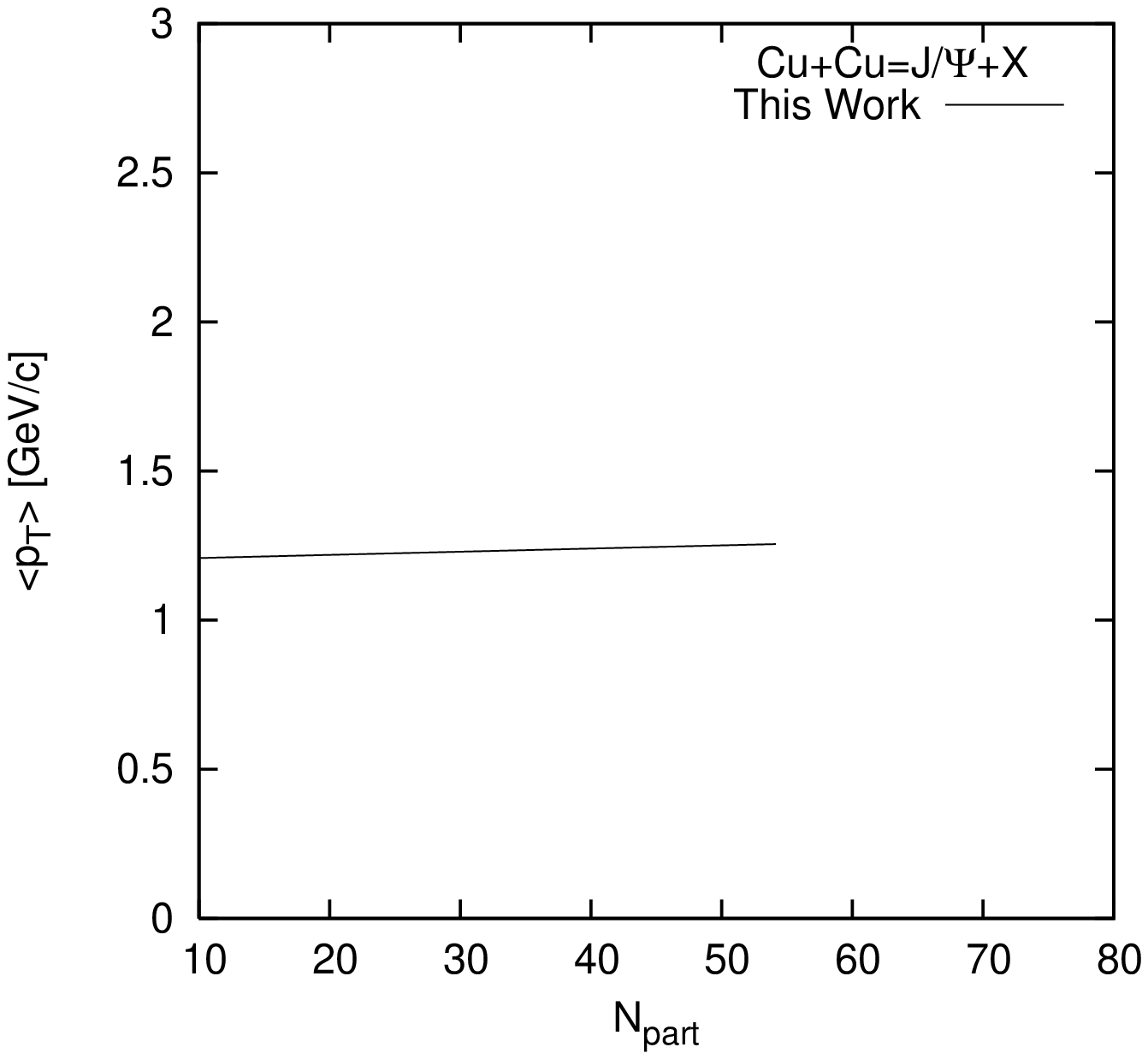}
\setcaptionwidth{2.6in}
\end{minipage}}%
\subfigure[]{
\begin{minipage}{0.5\textwidth}
\centering
 \includegraphics[width=2.5in]{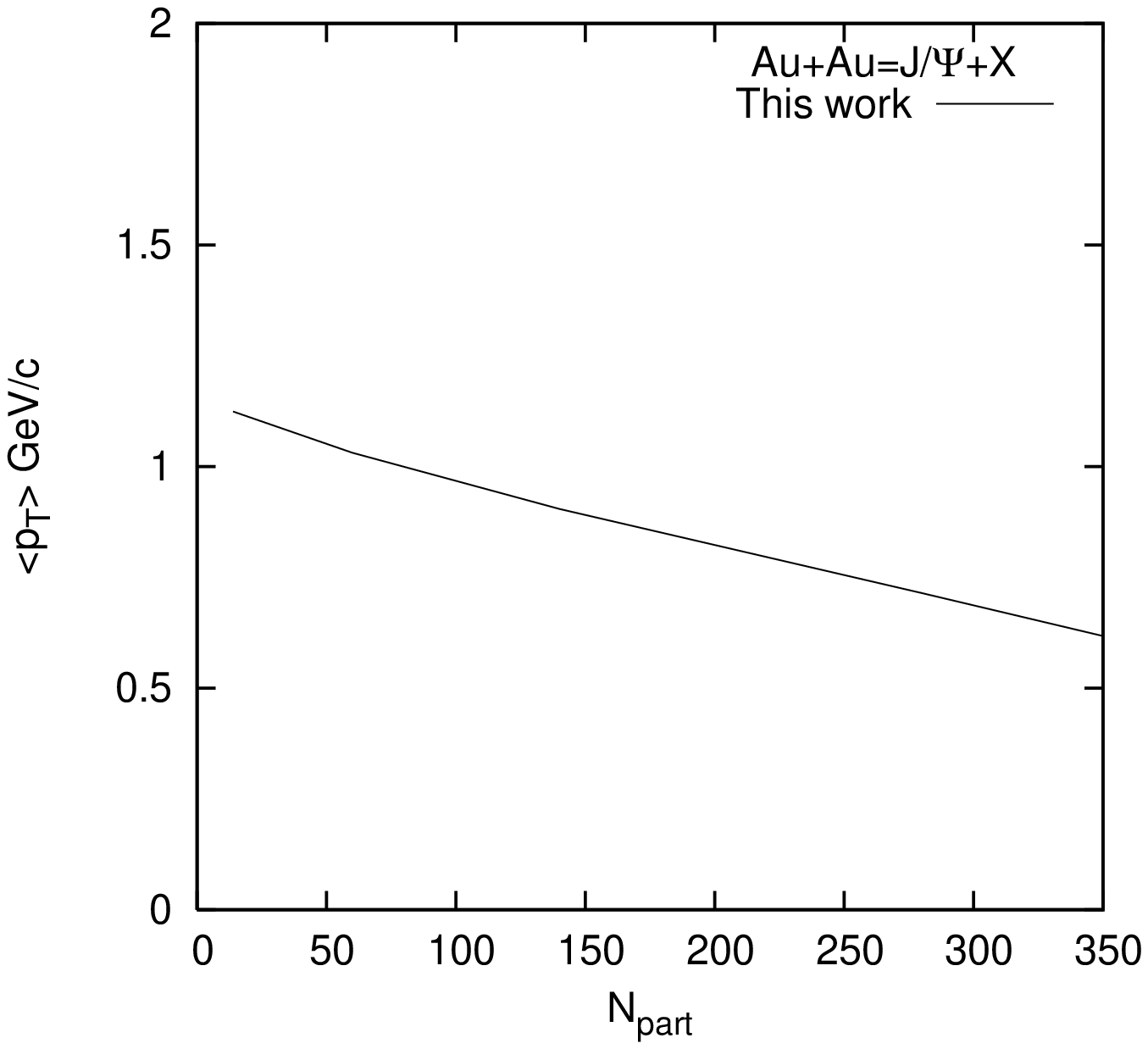}
 \end{minipage}}%
\caption{{\small $N_{part}$ vs.calculated results on $<p_T>^{J/\Psi}$ based on SCM for (a) $Cu+Cu$ and for (b) $Au+Au$.
The calculations are done on the basis of eqn.(4), Table 1 and Table 2. No experimental data on $<p_T>^{J/\Psi}$
are so far available. Calculated results are presented here in a predictive vein.}  }
\end{figure}
\begin{figure}
\subfigure[]{
\begin{minipage}{.5\textwidth}
\centering
\includegraphics[width=2.5in]{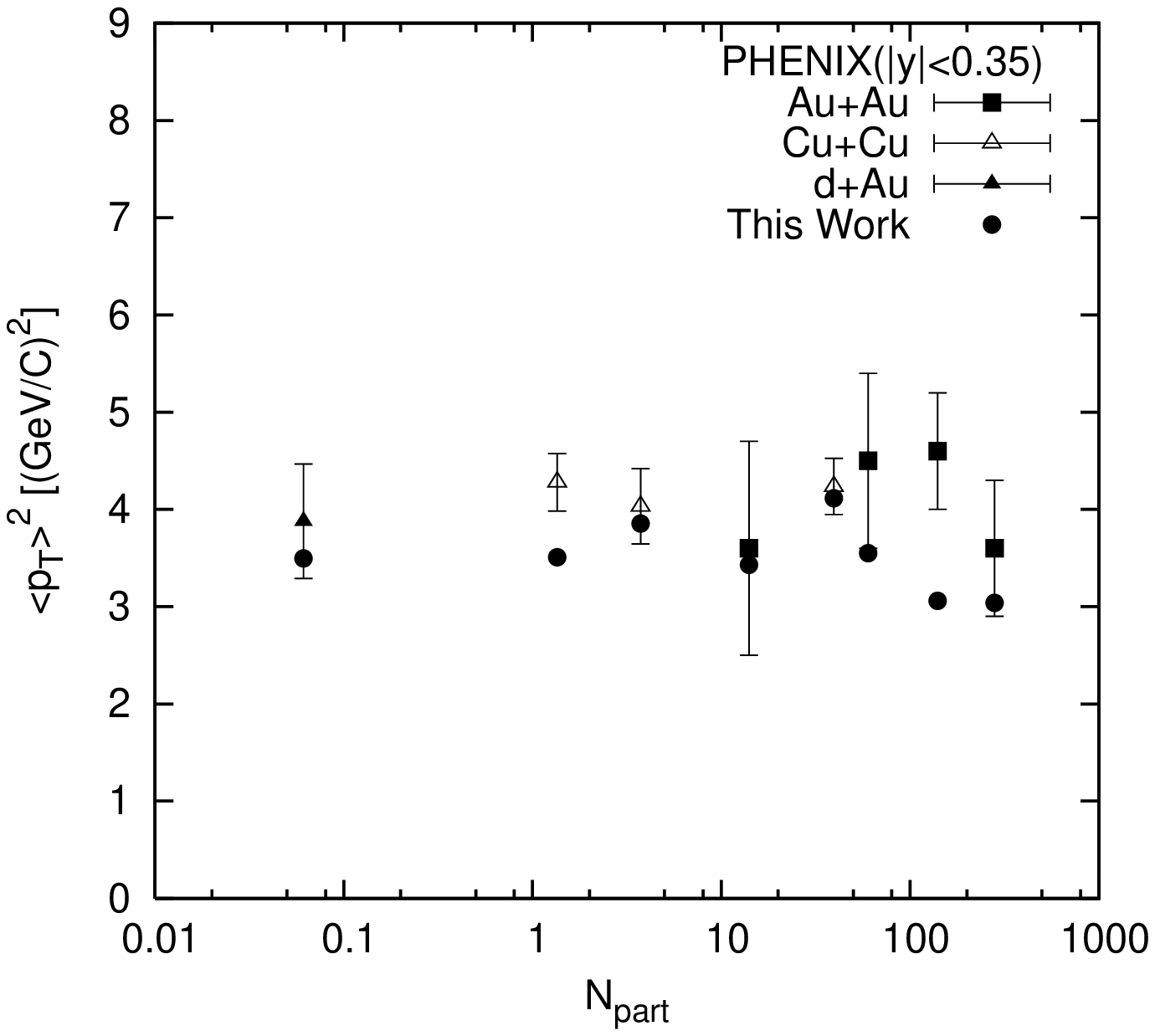}
\setcaptionwidth{2.6in}
\end{minipage}}%
\subfigure[]{
\begin{minipage}{0.5\textwidth}
\centering
 \includegraphics[width=2.5in]{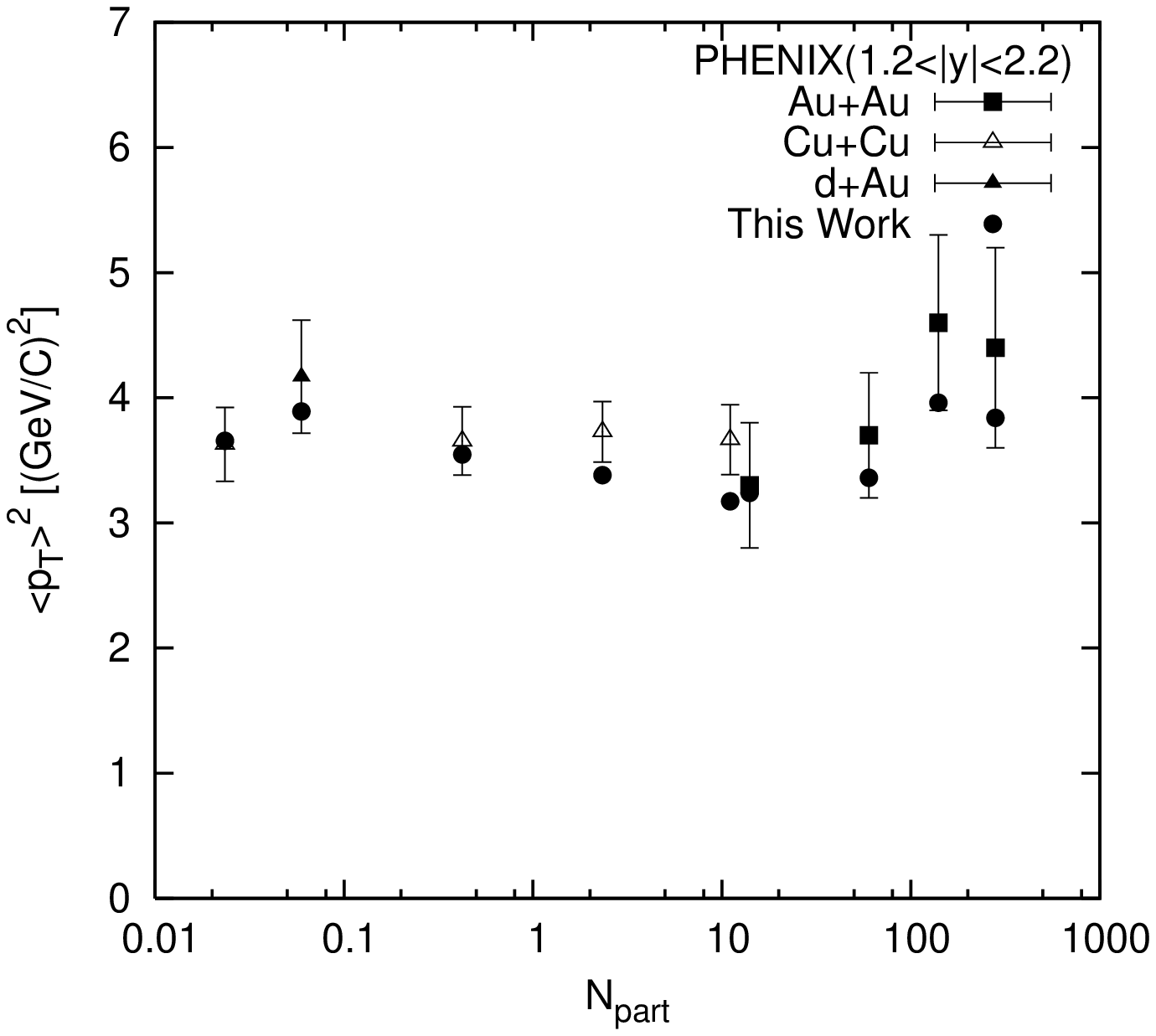}
 \end{minipage}}%
\caption{{\small Plots on $<p_T^2>$ versus $N_{part}$ for $J/\Psi$ production in $d+Au$, $Cu+Cu$ and $Au+Au$ collision in
(a) mid and (b) forward rapidities. The data points are taken from  \cite{adare4}, \cite{adare2}and \cite{adare3} and the
hollow circles in the Figure show the SCM-based results.}  }
\end{figure}
\begin{figure}
\subfigure[]{
\begin{minipage}{.5\textwidth}
\centering
\includegraphics[width=2.5in]{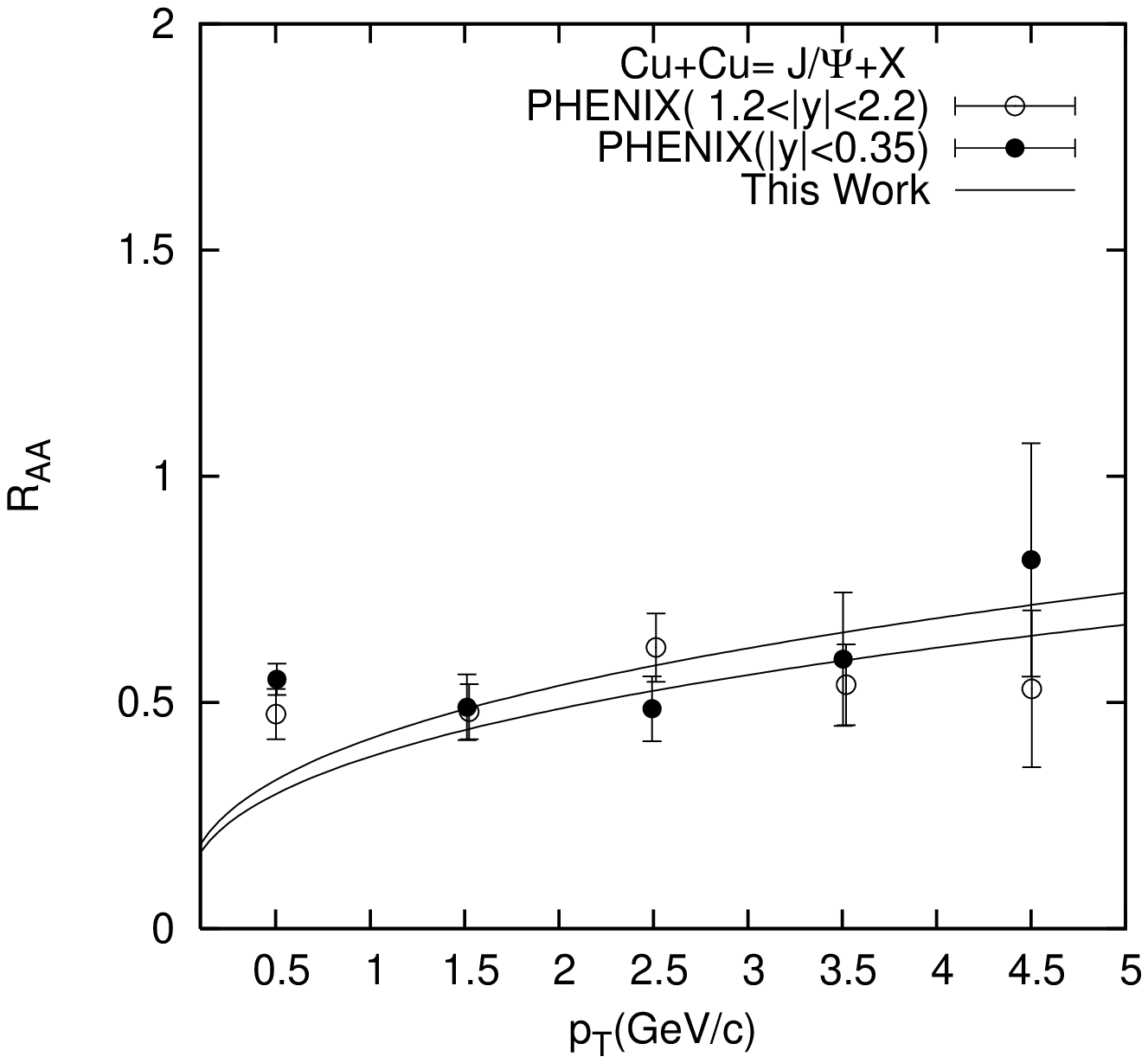}
\setcaptionwidth{2.6in}
\end{minipage}}%
\subfigure[]{
\begin{minipage}{0.5\textwidth}
\centering
 \includegraphics[width=2.5in]{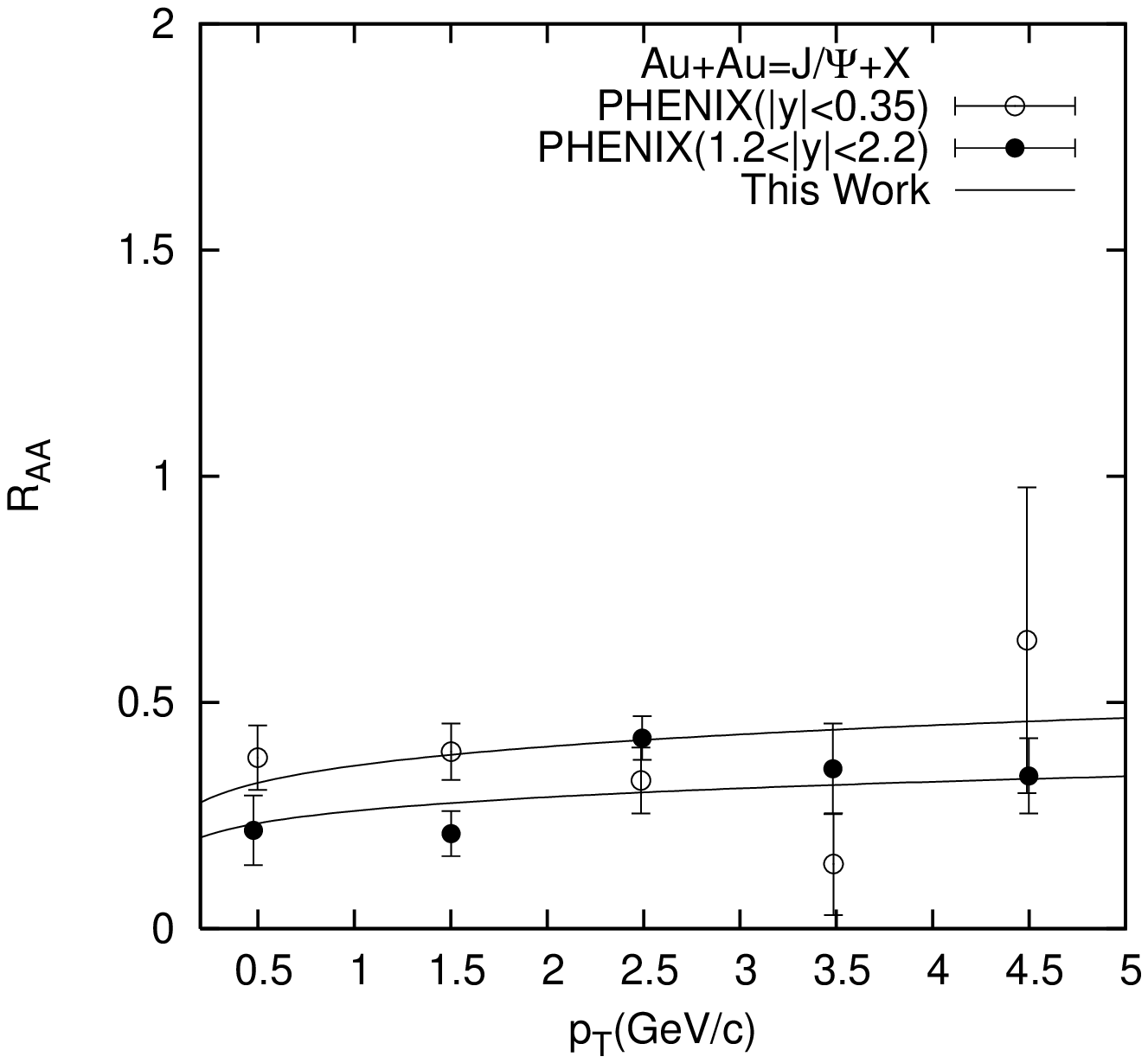}
 \end{minipage}}%
\caption{{\small Plot of $R_{AA}$ vs.
$p_T$-values for $0-20\%$ centrality in forward and
mid-rapidities. The data points for $Cu+Cu$ collisions (Fig. 8a) are taken from  Ref.\cite{adare2} while those for $Au+Au$ collisions (Fig. 8b) are taken from Ref. \cite{adare3} .
The solid lines show the SCM-based results.}  }
\end{figure}
\begin{figure}
\centering
\includegraphics[width=2.5in]{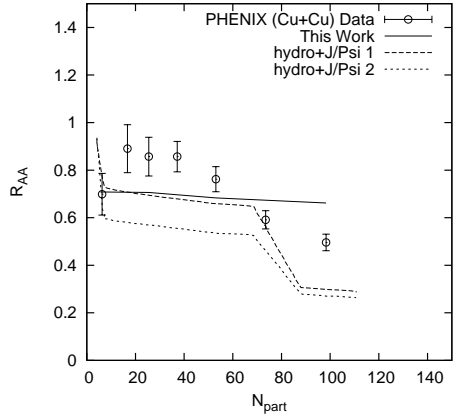}
 \caption{\small Nature of $N_{part}$-dependence of the nuclear modification
factor (NMF) denoted by
$R_{AA}$ plots for $Cu+Cu$ collisions at forward rapidity and at $\sqrt s_{NN} =200
GeV$. The data are taken from  Reference \cite{adare2}. The
SCM-based results are shown by the solid curved line in the
Figure.The dashed curves show the predictions of the hydrodynamical model (`hydro+$J/\Psi$') with (hydro+J/Psi1)and without (hydro+J/Psi2) nuclear absorption \cite{akc}.}
\end{figure}
\begin{figure}
\subfigure[]{
\begin{minipage}{.5\textwidth}
\centering
\includegraphics[width=2.5in]{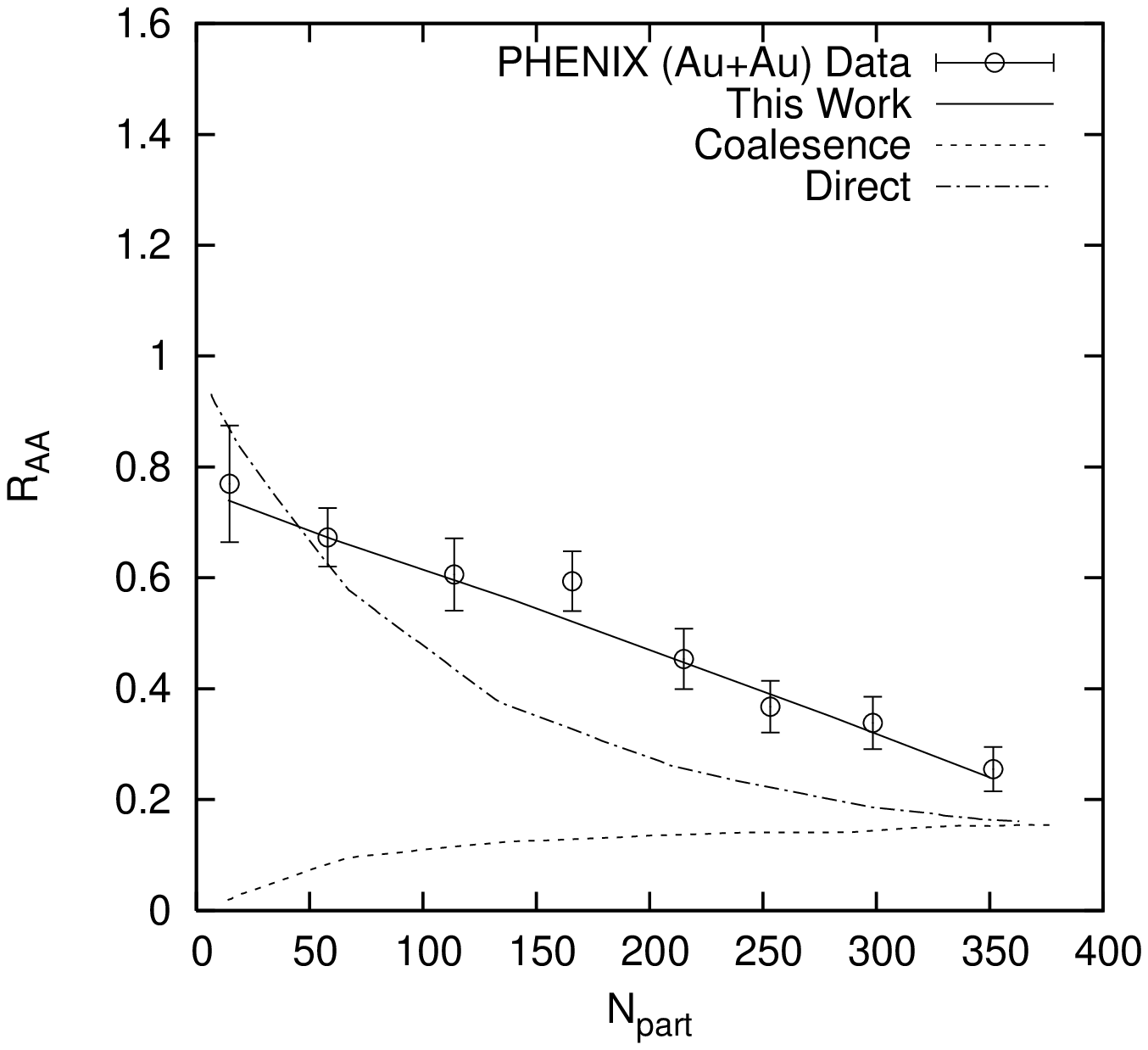}
\setcaptionwidth{2.6in}
\end{minipage}}%
\subfigure[]{
\begin{minipage}{0.5\textwidth}
\centering
 \includegraphics[width=2.5in]{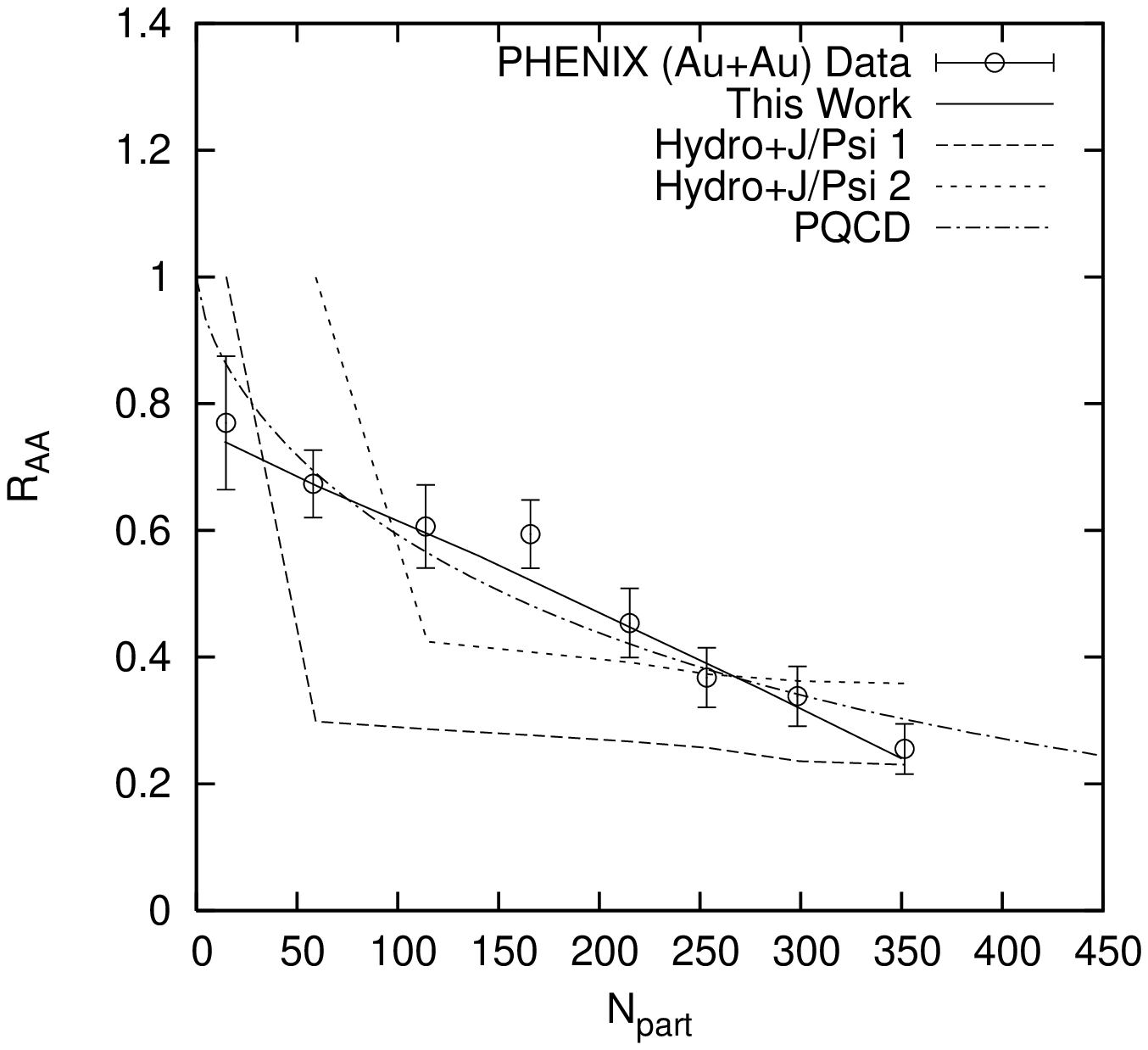}
 \end{minipage}}%
\caption{{\small Nature of $N_{part}$-dependence of the nuclear modification
factor (NMF) denoted by
$R_{AA}$ plots for $Au+Au$ collisions at forward rapidity and at $\sqrt s_{NN} =200
GeV$. The data are taken from \cite{adare3}, \cite{zhao}. The
SCM-based result is compared with other theoretical results (a) Coalescence and Direct approaches \cite{zhao} and (b)`hydro+$J/\Psi$' for $T_{J/\Psi}/T_C$= 1.2 and 1.4(hydro+J/Psi1, hydro+J/Psi2 respectively) \cite{akc}, and PQCD from Ref. \cite{vitev} in the
Figures.}  }
\end{figure}
\begin{figure}
\centering
\includegraphics[width=2.5in]{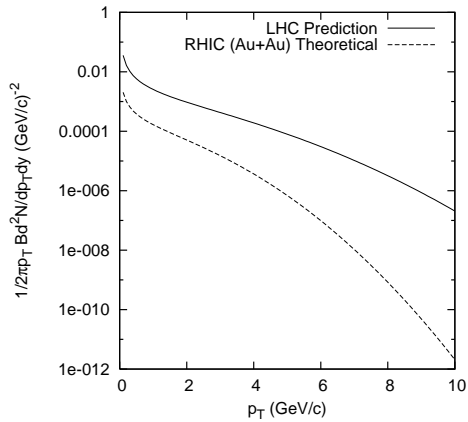}
\caption{\small SCM-based predicted plot on $J/\Psi$ production in the most central $Pb+Pb$ collisions
  at $\sqrt{s_{NN}}=5500$ GeV i.e. at Large Hadron Collider (LHC) energy is shown by the solid curve.
The dashed curve in the Figure represents our SCM-based plot for the most central collisions in $Au+Au$ collisions at
RHIC $\sqrt{s_{NN}}=200$ GeV energy.}
\end{figure}
\end{document}